\documentclass[twocolumn, twocolappendix, trackchanges]{aastex631}

\usepackage{threeparttable}
\usepackage{appendix}
\usepackage{natbib}
\usepackage{color}
\usepackage{mathptmx}
\usepackage{amsmath}
\usepackage{url}
\usepackage{multirow}
\usepackage{verbatim}
\usepackage[utf8]{inputenc}
\usepackage{bm}
\usepackage{gensymb}
\usepackage{booktabs}
\usepackage{soul}

\begin{document}

\shorttitle{From Dense Gas Clouds to Supermassive Black Hole Seeds}
\shortauthors{Chung et al.}

\title{From Dense Gas Clouds to Supermassive Black Hole Seeds: Hybrid Hydro/Direct $N$-body Simulations of Runaway Collision-driven Intermediate-mass Black Hole Formation}

\correspondingauthor{Ji-hoon Kim}
\email{mornkr@snu.ac.kr}

\author[0009-0002-3230-8205]{Eunwoo Chung}
\affiliation{Center for Theoretical Physics, Department of Physics and Astronomy, Seoul National University, Seoul 08826, Republic of Korea}

\author[0000-0003-3977-1761]{Yongseok Jo}
\affiliation{Center for Interdisciplinary Exploration and Research in Astrophysics (CIERA) and Department of Physics and Astronomy, Northwestern University, Evanston, IL 60208, USA}
\affiliation{NSF-Simons AI Institute for the Sky (SkAI), 172 E. Chestnut St., Chicago, IL 60611, USA}

\author[0000-0003-4464-1160]{Ji-hoon Kim}
\affiliation{Center for Theoretical Physics, Department of Physics and Astronomy, Seoul National University, Seoul 08826, Republic of Korea}
\affiliation{Institute for Data Innovation in Science, Seoul National University, Seoul 08826, Republic of Korea}
\affiliation{Seoul National University Astronomy Research Center, Seoul 08826, Republic of Korea}

\author[0000-0002-9144-1383]{Minyong Jung}
\affiliation{Center for Theoretical Physics, Department of Physics and Astronomy, Seoul National University, Seoul 08826, Republic of Korea}

\author{Oh-Kyoung Kwon}
\affiliation{National Supercomputing Center, Korea Institute of Science and Technology Information, Daejeon 34141, Republic of Korea}

\begin{abstract}

A population of dense stellar systems at high redshift has recently been uncovered by the JWST.
To investigate the formation of supermassive black hole (SMBH) seeds in these dense environments without invoking any \textit{ad hoc} seeding mechanisms, we present star cluster-scale simulations performed with an updated version of the hydrodynamics code \texttt{Enzo-Abyss}, which self-consistently integrates the gravity using a direct $N$-body method coupled with stellar evolution.
By modeling initially dense, metal-poor gas clouds with varying turbulence, we consistently find the formation of dense clusters resembling early-stage nuclear star clusters (NSCs), as well as the formation of very massive stars (VMSs) ranging from $343\;\mathrm{M_\odot}$ to $5108\;\mathrm{M_\odot}$ via runaway collisions, irrespective of stellar wind feedback strength.
Following the direct collapse of these VMSs, the resulting intermediate-mass black holes (IMBHs) grow through Eddington-limited gas accretion and tidal disruption events (TDEs).
In our most optimistic model, we find a mass accretion rate of $1.64\times10^{-4}\;\mathrm{M_\odot\;yr^{-1}}$, with TDEs contributing $23\%$ of the total accretion over $\sim10\;\mathrm{Myr}$.
Assuming a steady gas supply into the NSC driven by rapid structural assembly in the high-redshift environment, together with a constant TDE rate, we project that an IMBH with an initial mass of $6747\;\mathrm{M_\odot}$ at the center of the NSC can grow to $\sim62000\;\mathrm{M_\odot}$ within $100\;\mathrm{Myr}$ of its formation.
Our numerical study, conducted within a single self-consistent framework that incorporates the essential physical processes, suggests that VMSs can form in dense gas clouds, collapse into IMBHs, and subsequently provide viable seeds for the SMBHs observed at high redshift.

\end{abstract}



\hspace{1mm}

 \section{Introduction} \label{sec:intro}

The formation of supermassive black holes (SMBHs) could, in principle, be constrained by linking the massive populations observed in the local Universe to their high-redshift progenitors. 
Nevertheless, the origin of these objects remains one of the most significant open questions in astrophysics \citep{Volonteri10, Inayoshi20}.
The challenge has become even more acute with the advent of the James Webb Space Telescope (JWST), which has revealed a population of massive black holes already in place in the early Universe \citep{Goulding23, Larson23, Harikane23, Maiolino24, Kovacs24, Napolitano25}.
The presence of such massive objects merely hundreds of millions of years after the Big Bang places severe constraints on their formation theories, demanding either rapid growth rates \citep{Yoo04, Volonteri05} or massive initial seeds \citep{Lodato06}.

To explain the existence of SMBHs observed in the early Universe, several formation pathways have been proposed, including the direct collapse of primordial gas clouds, the remnants of Population III (Pop III) stars, and gravitational runaway collisions in dense star clusters.
Each scenario presents distinct advantages and theoretical challenges.
The direct collapse scenario predicts the formation of massive seeds ($M_\mathrm{BH} \gtrsim 10^5\;\mathrm{M_\odot}$) \citep{Loeb94, Eisenstein95, Begelman06}; 
however, preventing the fragmentation of the massive gas clouds into stars requires specific environmental conditions (e.g., strong Lyman-Werner radiation) that may be rare. 
Pop III stars are expected to be ubiquitous, providing a high occupation fraction of seeds \citep{Madau01, Volonteri03}. 
Yet, their remnants are relatively light ($\sim100\;\mathrm{M_\odot}$), necessitating sustained super-Eddington accretion to bridge the gap to the supermassive regime observed at high redshifts. 
Gravitational runaway collisions offer a mechanism that can function across a broad range of redshifts \citep{Sanders70, Lee87, Quinlan90}. 
The primary constraint for this scenario, however, is that extreme central densities are required to promote successive stellar mergers during the short lifetime ($\lesssim 5\;\mathrm{Myr}$) of the resulting very massive stars (VMSs).
While the existence of VMSs at high redshift has long been a subject of theoretical debate, recent observations \citep[e.g.,][]{Marques-Chaves26} provide evidence that VMSs may be significantly more overabundant in the early Universe than previously anticipated.

Remarkably, recent observations by the JWST have unveiled the existence of young, dense stellar systems at $z = 6-10$, interpreted as proto-globular clusters \citep{Vanzella23, Mowla24, Bradac25, Messa25a, Fujimoto25, Abdurrouf25}.
In particular, the Cosmic Gems arc \citep{Adamo24, Bradley25, Vanzella25} offers one of the most intriguing cases.
Through gravitational lensing, this system reveals five star clusters at $z = 9.6$ with ages younger than $50\;\mathrm{Myr}$, sizes of $\sim 1\;\mathrm{pc}$, and masses of $\sim 10^6\;\mathrm{M}_\odot$.
These properties yield stellar surface densities of approximately $10^5\;\mathrm{M}_\odot\;\mathrm{pc}^{-2}$, values comparable to those of massive nuclear star clusters (NSCs) found in the centers of galaxies \citep{Neumayer20}.

In these dense environments, the VMSs can grow to masses exceeding hundreds or even thousands of solar masses \citep{Portegies02, Portegies04}.
Crucially, provided these stars are sufficiently massive to avoid complete disruption via pair-instability supernovae (PISNe), they are predicted to collapse directly into intermediate-mass black holes (IMBHs) with little to no mass loss \citep{Heger02, Spera17}.
Thus, the recent discovery of dense star clusters at high redshift highlights the viability of this scenario, offering a robust and physically motivated pathway for seeding massive black holes that bypasses the growth bottlenecks associated with lighter stellar remnants.

Following their formation, IMBHs are expected to grow primarily via gas accretion.
In the gas-rich cores of young NSCs at high redshift, the central black hole is embedded in a dense reservoir, allowing for sustained accretion that may approach or even exceed the Eddington limit \citep{Inayoshi16, Wu25}.
Simultaneously, extreme stellar densities resulting from dynamical friction and mass segregation significantly increase the probability of close encounters between the IMBH and surrounding stars, triggering tidal disruption events \citep[TDEs; ][]{Rees88, Strubbe09, Rizzuto23}.
However, despite their potential to accelerate black hole growth, the contribution of TDEs remains difficult to capture self-consistently in hydrodynamical studies.
For instance, while \citet{Lee23} investigated the evolution of massive black hole seeds inside dense NSCs, the computational expense of resolving individual stars forced them to rely on analytical models for stellar density profiles and TDE rates.

To investigate the complete evolutionary sequence --- from the formation of dense star clusters and the onset of runaway collisions to the birth and subsequent growth of IMBHs --- in a self-consistent manner, we introduce an updated version of the hybrid hydro/$N$-body framework, \texttt{Enzo-Abyss}.
The development of such multi-physics frameworks is a growing necessity in modern astrophysics, as seen in recent efforts \citep[e.g.,][]{Wall20, Cournoyer-Cloutier21, Cournoyer-Cloutier25, Grudic21, Farias24, Bernard25} to couple high-precision stellar dynamics with hydrodynamics.
These studies have pushed the boundaries of resolving star cluster formation and evolution within their natal gaseous environments.

\begin{figure*}[t]
    \vspace{-1mm}
    \centering
    \includegraphics[width=\textwidth]{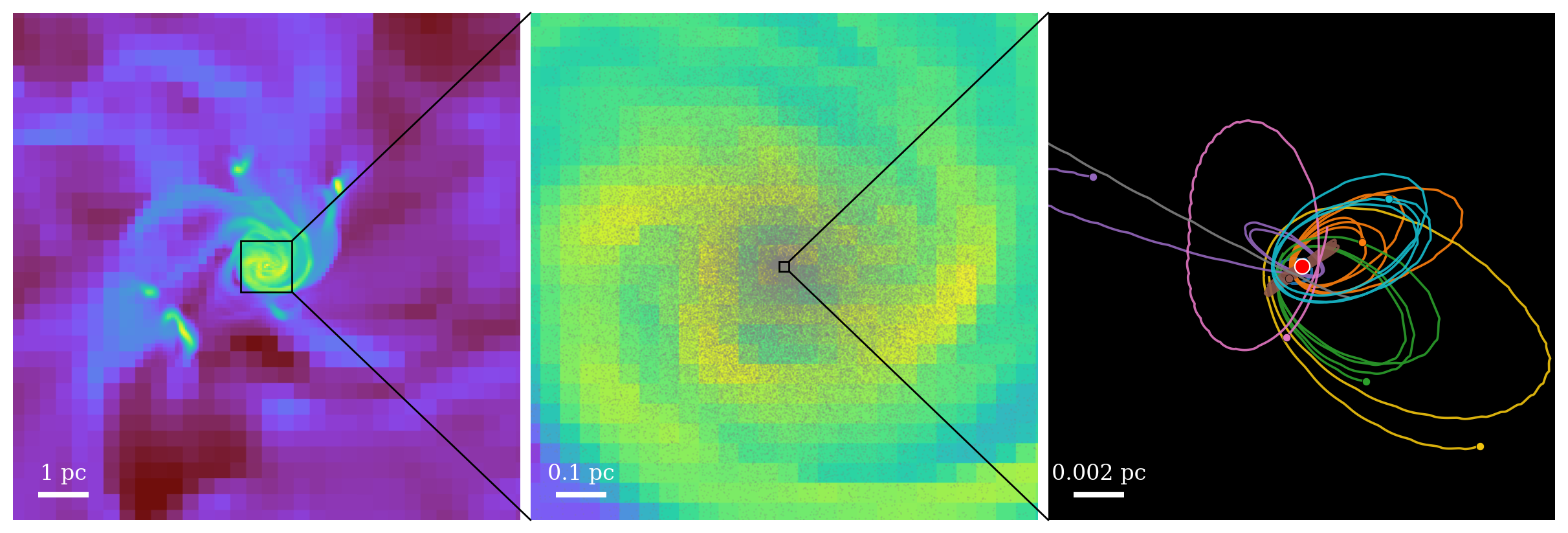}
    \caption{
    Multi-scale overview of the \texttt{Enzo-Abyss} simulation for the model with $\alpha_\mathrm{vir} = 1.5$ and $v_\mathrm{w} = 3000\;\mathrm{km\;s^{-1}}$.
    \textit{Left}: Projected gas density within the central $10\;\mathrm{pc}$.
    \textit{Middle}: A zoom-in of the central $1\;\mathrm{pc}$ region, showing the gas density overlaid with the stellar distribution (\textit{gray points}).
    \textit{Right}: Trajectories of the 10 stars nearest to the central IMBH (\textit{red circle}; see Section \ref{subsec:IMBH}) over $1000\;\mathrm{yr}$, illustrating orbits computed on the fly with the exact $N$-body solver inside \texttt{Enzo-Abyss} (see Section \ref{subsec:abyss}).
    Note that the non-smooth appearance of some orbits in the right panel arises from the second-order extrapolation used to synchronize individual particle timesteps for visualization and the frequent stellar perturbations in the dense core; the underlying N-body integration is fourth-order accurate.
    The final position of each star is marked by a circle.
    All panels are centered on the position of the IMBH.
    TDEs are triggered when a star passes within the tidal radius of the IMBH; half of the stellar mass is accreted onto the IMBH and the other half is returned to the surrounding gas (Section \ref{subsubsec:TDA}).
    }
    \label{zoom-in}
    \vspace{2mm}
\end{figure*}

Previously, \citet{Jo24} integrated the direct $N$-body code \texttt{NBODY6++GPU} \citep{Wang15} into the hydrodynamics code \texttt{Enzo} to simulate star cluster evolution within galaxies.
However, that implementation was limited by its inability to resolve the physics of individual stars (i.e., the lack of appropriate star formation and feedback prescriptions).
Furthermore, coupling \texttt{NBODY6++GPU} with the gas dynamics of \texttt{Enzo} proved technically challenging, as the $N$-body code was originally designed for standalone use.
To overcome these hurdles, we have developed a new direct $N$-body code, \texttt{Abyss} \citep{Jo26}.
This code is seamlessly coupled with modern external libraries for small-scale physics --- such as \texttt{SDAR} \citep{Wang20} for accurate few-body dynamics and \texttt{SEVN} \citep{Iorio23} for stellar evolution --- and is fully integrated within the \texttt{Enzo} platform.
By implementing robust individual star physics and improved integration algorithms, we can now resolve individual stellar dynamics with high-precision gravity, eliminating the need for gravitational softening.
The present study, conducted within this single self-consistent framework that incorporates all relevant physical processes, demonstrates that VMSs can form in dense gas clouds, collapse into IMBHs, and subsequently provide viable seeds for the SMBHs observed at high redshift.

The paper is structured as follows.
In Section \ref{sec:numerical}, we describe the \texttt{Enzo-Abyss} framework and provide a detailed explanation of its components.
In Section \ref{sec:simulations}, we outline the adopted initial conditions and the physical models governing stellar and black hole evolution.
The main results regarding VMS/IMBH formation and growth are presented in Section \ref{sec:results}, followed by a further discussion in Section \ref{sec:discussion}.
Finally, we summarize our key findings and conclude in Section \ref{sec:conclusion}.

\hspace{1mm}

\section{Numerical Methods} \label{sec:numerical}

Figure \ref{zoom-in} presents a multi-scale overview of the \texttt{Enzo-Abyss} simulation. 
By incorporating all the numerical components needed to describe the complete evolutionary sequence of SMBH seeds --- from the formation of dense star clusters to the onset of runaway collisions and the growth of IMBHs --- 
this framework allows us to accurately track these physical processes across a vast range of spatial scales.

\subsection{\textup{\texttt{Enzo}}: Adaptive Mesh Refinement Hydrodynamics Code} \label{subsec:enzo}

\texttt{Enzo} is a block-structured Eulerian adaptive mesh refinement code, developed for a variety of astrophysical problems \citep{Bryan14}. 
In this study, the 3rd order accurate piecewise parabolic method (PPM) \citep{Colella84, Bryan95} with a Harten-Lax-van Leer with Contact (HLLC) Riemann solver \citep{Toro94} is employed as the hydrodynamic solver.
\texttt{Enzo} uses the Particle Mesh (PM) gravity solver to deal with the gravitational interaction between the baryons and collisionless particles, such as dark matter.
The motion of particles are integrated at every timestep via a drift-kick-drift algorithm \citep{Hockney88}, which is accurate to the second order.
\texttt{Enzo} also has the well-tested subgrid prescriptions, including star formation, stellar feedback, massive black hole accretion, and active galactic nuclei (AGN) feedback.  

\subsection{\textup{\texttt{Abyss}}: Direct $N$-body Code} \label{subsec:abyss}

\texttt{Abyss} is a direct $N$-body code employing the fourth order Hermite integrator and the block time step method \citep{Jo26}.
\texttt{Abyss} can be used as a standalone code, but it is integrated into \texttt{Enzo} to model collisional gravitational dynamics as in \texttt{Enzo-N} framework \citep{Jo24}.
In this framework, particles within a user-defined region are integrated via \texttt{Abyss} without gravitational softening.
In this study, we defined this region to encompass the entire simulation box;
consequently, the motions of all stars and black holes are integrated without softening.
By adopting the Ahmad-Cohen neighbor scheme \citep{Ahmad73}, the total force acting on a particle in \texttt{Abyss} is divided into two components:  an irregular force and a regular force.
The irregular force, exerted by nearby neighbor particles\footnote{In this study, the neighbor list of a particle is dynamically updated to maintain a neighbor count of 50.
}, is updated frequently on an irregular timestep and is parallelized using MPI.
In contrast, the regular force from distant particles changes more smoothly, is updated on a less frequent regular timestep, and is parallelized on GPUs\footnote{While the standard CPU version of \texttt{Abyss} computes regular forces across multiple nodes, the version employed in this study accelerates these calculations on GPUs within a single node. A multi-node GPU implementation is currently under development (Chung et al. in prep.).}.
While the code's architecture is largely inspired by the standard direct $N$-body code, \texttt{NBODY6++GPU} \citep{Wang15}, we introduce a novel queue scheduling system for dynamic load balancing:
a root processor distributes tasks to worker processors asynchronously as they become available, minimizing processor idling and maximizing computational efficiency.
Furthermore, a shared-memory interface utilizing MPI-3.0 has been implemented to reduce the communication overhead when distributing particle data.
More technical details will be presented in Chung et al. (in prep.).

\subsection{\textup{\texttt{SDAR}}: Slow-Down Algorithmic Regularization} \label{subsec:sdar}

In the dense core of an $N$-body system, hard binaries and highly eccentric orbits can frequently occur.
However, the Newtonian gravity has the singularity, and the Hermite method becomes largely inaccurate or time-consuming at the pericenter.
To resolve the issue, a special treatment for close encounters or compact few-body systems is required.
The regularization technique can be used to avoid the singularity by transforming the equation of motion \citep{Kustaanheimo65, Mikkola99, Preto99}.

Here, we implement the slow-down algorithmic regularization \citep[SDAR;][]{Wang20} integrator into \texttt{Abyss} to accurately evolve the orbit of compact few-body groups.
The widely used \texttt{SDAR} library\footnote{Here, \texttt{SDAR} denotes the software library, while SDAR refers to the algorithm itself. \texttt{SDAR} is publicly available at \url{https://github.com/lwang-astro/SDAR}.} combines the algorithmic regularization and the slow-down technique, which artificially slows down the orbital motion of weakly perturbed binaries \citep{Mikkola96}.
Thus the effect of external perturbation is enhanced, and the corresponding orbital motion can be approximated to the secular evolution \citep{Wang20}.
While it loses the phase information of orbits, it can efficiently reduce the computational cost, so various modern $N$-body codes, such as \texttt{PETAR} \citep{Wang20Petar} and \texttt{BIFROST} \citep{Rantala21, Rantala23}, use \texttt{SDAR} to integrate few-body dynamics.

An orbit integration of a group of particles is performed by \texttt{SDAR} when particles in the neighbor list approach closer than a threshold distance, which is set to $10^{-4}\;\mathrm{pc}$ in this study.
For the force calculation in \texttt{Abyss}, such a group is treated as a single center-of-mass particle.
Few-body systems involving three or more particles can form through close encounters, including those involving existing center-of-mass particles.
In our simulations, we did not encounter systems composed of more than four particles. 
A system is dissolved into individual particles if the separation between any pair of particles within the binary tree --- automatically constructed by $\texttt{SDAR}$ --- exceeds the threshold distance and the particles are receding from each other.
This dissolution strategy is adopted to break up systems with more than two particles early, as long-lasting many-body systems can significantly increase the computational cost.

\subsection{\textup{\texttt{SEVN}}: Stellar EVolution for $N$-body} \label{subsec:sevn}

In order to consider stellar feedback from individual stars, it is necessary to trace detailed stellar evolution from the zero-age main sequence (ZAMS) to the final evolutionary phase.
Various packages are developed for stellar evolution in direct $N$-body codes, e.g., \texttt{SSE/BSE} \citep{Hurley00, Hurley02}, \texttt{MOBSE} \citep{Giacobbo18}, \texttt{BSEEMP} \citep{Tanikawa20}, and \texttt{SEBA} \citep{Portegies96, Toonen12}.
Among those codes, we choose \texttt{SEVN}\footnote{\texttt{SEVN} is publicly availble at \url{https://gitlab.com/sevncodes/sevn}.} \citep{Iorio23} to be implemented into \texttt{Abyss}.
\texttt{SEVN} is a rapid population synthesis code, which is designed to calculate stellar evolution through interpolation of precomputed stellar tracks \citep{Iorio23, Spera15, Spera17}.
\texttt{SEVN} uses the adaptive time-step to prevent drastic change of stellar properties.

\texttt{SEVN} offers two types of stellar tracks calculated from \texttt{PARSEC} \citep{Bressan12, Costa19, Costa21, Nguyen22} and \texttt{MIST} \citep{Choi16}.
We choose \texttt{SEVNtracks\_parsec\_ov04\_AGB} since these tracks deal with the largest range of stellar mass ($2.2\;\mathrm{M_\odot}\leq M_\star \leq 600\;\mathrm{M_\odot}$), including VMS range.\footnote{The tracks are calculated from the overshooting parameter of $\lambda_\mathrm{ov} = 0.4$.  See \citet{Iorio23} for details.}
For stars below this range ($M_\star < 2.2\;\mathrm{M_\odot}$), we ignore stellar evolution, as their lifetimes far exceed our simulation timescale of $\sim10\;\mathrm{Myr}$.
Conversely, runaway mergers can produce VMSs exceeding the upper limit of the tracks ($M_\star > 600\;\mathrm{M_\odot}$).
For these objects, we extrapolate the $600\;\mathrm{M_\odot}$ track: we assume the star retains the same lifetime and evolutionary phase as a $600\;\mathrm{M_\odot}$, but we scale the mass loss rate by a factor of $M_\star/600\;\mathrm{M_\odot}$, following \citet{Rantala24}.
The evolution of such massive stars carries significant theoretical uncertainty.
For instance, \citet{Vergara25} incorporate stellar rejuvenation after stellar mergers, resulting in extended lifetimes for VMSs. 
In this study, we restrict our analysis to the simplified extrapolation described above, deferring further improvements to future work.

\texttt{SEVN} also offers binary stellar evolution which is calculated through analytic and semi-analytic methods, but it is not considered in this study.
Even though multiple stars are identified as a few-body system, we evolve them as single stars.
Binary stellar evolution with primordial binary formation will be considered in future works.


\section{Simulations} \label{sec:simulations}

\subsection{Refinement Strategy and Initial Conditions} \label{subsec:ic}

Our computational domain is a $(40\;\mathrm{pc})^3$ box with a root grid of $128^3$ cells, corresponding to an initial spatial resolution of $0.31\;\mathrm{pc}$.
We apply a Jeans length refinement criterion, ensuring that the local Jeans length is resolved by at least 4 cells, as suggested by \citet{Truelove97}.
Additionally, a cell is refined into 8 child cells if the cumulative particle mass it contains exceeds $600\;\mathrm{M}_\odot$.
This criterion ensures that the hydrodynamical grid maintains high resolution in regions where the $N$-body dynamics drive high stellar densities.
The maximum refinement level ($l_\mathrm{max}$) is set to be 3, achieving the finest cell size of $\Delta x = 0.04\;\mathrm{pc}$.

We initialize a dense gas cloud of metallicity $Z = 0.02\;Z_\odot$\footnote{We adopt a solar metallicity value of $Z_\odot = 0.02041$.} with a density profile given by
\begin{equation}
    \rho(r) = \rho_\mathrm{c} \left( 1 + \frac{r^2}{a^2} \right) ^ {-2},
    \label{eq:fiducial profile}
\end{equation}
where the central density is $\rho_\mathrm{c} = 1.29\times10^{-20}\;\mathrm{g}\;\mathrm{cm}^{-3}$ and the scale length is $a = 5\;\mathrm{pc}$.
These parameters result in a total gas mass of $M_\mathrm{gas} = 1.75\times 10^5\;\mathrm{M}_\odot$ within the simulation domain.
The initial density is chosen to be sufficiently high to promote bursty star formation and the formation of VMSs through mergers. 
The mean surface density within the central $5\;\mathrm{pc}$ is $550\;\mathrm{M_\odot\;pc^{-2}}$, which is in line or lower than the constant initial density profiles ranging from $1100$ to $3200\;\mathrm{M_\odot\;pc^{-2}}$ used by \citet{Fujii24}.
We explore the sensitivity of our results to the slope of the density profile by comparing this fiducial model with a Plummer profile in Section \ref{subsec:plummer}.
We apply a turbulent velocity field on the initial $128^3$ cells, characterized by a power spectrum of $v_\mathrm{k}^2 \propto k^{-4}$, consistent with previous hydrodynamical simulations of turbulent molecular clouds \citep[e.g.,][]{Bonnell03, Wang10, Fujii24}.
The strength of the initial turbulence is set to yield a specific virial parameter, defined as $\alpha_\mathrm{vir} = 2E_\mathrm{K} / |E_\mathrm{P}|$, where $E_\mathrm{K}$ is the kinetic energy and $E_\mathrm{P}$ is the gravitational potential energy.
Our fiducial models are initialized in virial equilibrium ($\alpha_\mathrm{vir} = 1.0$).
In addition, to reflect the highly turbulent conditions of the high-redshift Universe, we also explore super-virial models with $\alpha_\mathrm{vir} = 1.5$ and $3.0$.
Our simulation suite with varying $\alpha_\mathrm{vir}$ values is summarized in Table \ref{tab:VMS_mass}.

\subsection{Star Formation} \label{subsec:formation}

Since our simulations resolve individual stars, we employ the stochastic star formation algorithm of \citet{Goldbaum15, Goldbaum16}, incorporating the updates described by \citet{Emerick19}.
We define a cell to be star forming candidate, if it is at the maximum refinement level and satisfies the following criteria:
(1) the hydrogen number density exceeds the threshold $n_\mathrm{thres} = 2\times10^5\;\mathrm{cm}^{-3}$,
(2) the temperature is lower than the threshold $T_\mathrm{thres} = 30\;\mathrm{K}$,\footnote{To compute the cooling and heating rates, we employ the \texttt{GRACKLE} library \citep[][https://grackle.readthedocs.io]{Smith17}. \texttt{GRACKLE} provides tabulated metal cooling rates derived from \texttt{CLOUDY} \citep{Ferland13}, as well as the UV background photo-heating and photo-ionization rates based on \citet{Haardt12}. In this work, we employed the equilibrium cooling version of \texttt{GRACKLE}. We plan to implement a non-equilibrium primordial chemistry network in future studies with \texttt{AEOS} physics module (see Section \ref{subsec:future}).}
(3) the local velocity field is converging $\nabla\cdot v < 0$, and
(4) the gas mass in the cell is higher than the local Jeans mass $M > M_\mathrm{Jeans}$, where $M_\mathrm{Jeans}$ is approximately $12\;\mathrm{M}_\odot$ with the cell temperature $T = 30\;\mathrm{K}$.

In this study, we adopt a Kroupa initial mass function \citep[IMF;][]{Kroupa01} with a mass range from $M_\mathrm{IMF,\;min} = 0.08\;\mathrm{M}_\odot$ to $M_\mathrm{IMF,\;max} = 150\;\mathrm{M}_\odot$.
To sample stars from the IMF, the mass of star forming gas cell $M$ must be greater than at least $M_\mathrm{IMF,\;max}$, the upperbound of our chosen IMF.
This requirement could create unnaturally dense cells, as the Jeans mass $M_\mathrm{Jeans}$ is typically much lower than $M_\mathrm{IMF,\;max}$ in our simulation.
To prevent this problem, we gather the gas from surrounding 26 $(=3^3 -1)$ cells if a cell's density is greater than $n_\mathrm{thres}$.
Star formation is permitted only if the cumulative gas mass of the star-forming cell and its surroundings exceeds a threshold of $f_\mathrm{thres} M_\mathrm{IMF,\;max}$, where $f_\mathrm{thres}$ is a free parameter and chosen to be 2 in our study.\footnote{This approach is inline with \citet{Hirai21} that adopts a smoothed particle hydrodynamics (SPH) code and employs a similar star formation method.  They introduce the search radius and gather gas particles within the radius, if a star-forming gas particle mass is smaller than a randomly sampled stellar mass from the IMF.}

\begin{table}[t]
    \centering
    \caption{
    Simulation parameters ($\alpha_\mathrm{vir}$ and $v_\mathrm{w}$), resulting VMS masses ($M_\mathrm{VMS}$), total number of mergers onto the VMS progenitor ($N_\mathrm{merger}$), total number of mergers during the main sequence phase of the VMS progenitor ($N_\mathrm{merger,\;MS}$), and time elapsed from the first star formation to IMBH formation ($t_\mathrm{IMBH}$).
    }
    \label{tab:VMS_mass}
    \begin{tabular}{cc||cccc} 
        \hline\hline
        \boldmath{$\alpha_\mathrm{vir}$} & \boldmath{$v_{\text{w}}$} & \boldmath{$M_{\text{VMS}}$} & \boldmath{$N_{\text{merger}}$} & \boldmath{$N_{\text{merger, MS}}$} & \boldmath{$t_\mathrm{IMBH}$} \\
         & [km s$^{-1}$] & [$\mathrm{M_{\odot}}$] & & & [Myr] \\
        \hline
        1.0 & 500  & 5108 & 2543 & 1379 & 2.63 \\
        1.0 & 3000 & 4142 & 1625 & 1021 & 2.58 \\
        1.5 & 500  & 3950 & 1605 & 971 & 3.22 \\
        1.5 & 3000 & 3299 & 1535 & 911 & 2.82 \\
        3.0 & 500  & 519  & 141 & 80 & 2.94 \\
        3.0 & 3000 & 343  & 63 & 49 & 2.85 \\
        \hline
    \end{tabular}
\end{table}

Additionally, we adopt a stochastic star formation model \citep{Goldbaum15, Goldbaum16} designed to reproduce the local Schmidt law,
\begin{equation} \label{eq:schimidt law}
    \frac{d\rho_\star}{dt} = \varepsilon_\star\frac {\rho_\mathrm{gas}}{t_\mathrm{ff}},
\end{equation}
where $\rho_\star$ is the stellar density, $\rho_\mathrm{gas}$ is the gas density, $t_\mathrm{ff} = (3\pi/(32G\rho_\mathrm{gas}))^{1/2}$ is the local free-fall time, and $\varepsilon_\star$ is the star formation efficiency, which is fixed to 0.01 in this study.
In this scheme, a star formation event is triggered if a random number $P$, drawn uniformly from $[0, 1)$, satisfies $P \leq P_\star$.
The threshold probability $P_\star$ is derived by discretizing  Eq. \ref{eq:schimidt law} over the timestep $\Delta t$ and cell volume $\Delta x^3$:
\begin{equation} \label{eq:stochastic SF}
    P_\star = \varepsilon_\star \frac{\rho_\mathrm{gas} \Delta x^3}{M_\mathrm{IMF,\;max}} \frac{\Delta t}{t_\mathrm{ff}}.
\end{equation} 
When this condition is met, we generate stars by sampling the IMF until the cumulative mass exceeds $M_\mathrm{IMF,\;max}$, ensuring the IMF is fully populated without truncation bias.\footnote{In Eq.(\ref{eq:stochastic SF}), $M_\mathrm{IMF,\;max}=150\;\mathrm{M_\odot}$ acts as a fixed mass threshold. However, the actual total mass of the formed stars is allowed to slightly exceed this value to ensure individual stars are sampled without bias.}

To prevent unnatural stellar mergers immediately following formation in extremely dense environments, newly formed stars are randomly deposited within a cubic volume of width $0.156\;\mathrm{pc}$, centered on the host gas cell.
Stellar velocities are initialized from a Gaussian distribution centered on the bulk velocity of the gas, with a dispersion of $1\;\mathrm{km\;s^{-1}}$, consistent with observations \citep{Foster15}.
Although stellar positions and velocities are initially stochastic, they rapidly settle into a dynamically stable configuration under their self-gravity, calculated in \texttt{Abyss} using a direct $N$-body method without softening.
Following initialization, we apply corrections to ensure the local conservation of the center-of-mass position and total momentum of the gas-star system.
Primordial binary formation is not included in this work and all stars are initialized as single objects, though binaries may subsequently form through dynamical interactions.

\subsection{Stellar Feedback} \label{subsec:stellarfeedback}

We improve \texttt{Enzo-N} framework \citep{Jo24} by implementing stellar evolution code \texttt{SEVN} \citep{Iorio23} in \texttt{Abyss} to trace the evolution of individual stars.
By coupling \texttt{SEVN} with hydrodynamics code \texttt{Enzo} \citep{Bryan14}, feedback injections from individual stars in different evolutionary phases is modeled.
In this section, we describe in detail on how the stellar feedback process is treated.
Regarding chemical enrichment, we do not track individual chemical yields in this work; 
instead, we assume a constant metal yield fraction of 0.02 for both stellar winds and supernova events.

\subsubsection{Stellar Wind Feedback} \label{subsubsec:winds}

As one of pre-SN feedback sources, we implement stellar wind feedback.
For each \texttt{Enzo} timestep, stellar positions, velocities, and properties are evolved in \texttt{Abyss} and sent to \texttt{Enzo}.
If a star has stellar mass loss within an \texttt{Enzo} timestep, the corresponding mass and energy are injected into surrounding cells.
We assume complete thermalization of the wind's kinetic energy and inject the feedback as pure thermal energy.
Because our simulations reach a maximum resolution of $0.04\;\mathrm{pc}$, the feedback injection occurs on scales small enough to resolve the Sedov-Taylor phase \citep[see][]{Simpson15, Emerick19} and significantly alleviate the overcooling problem, which typically occurs at lower resolutions.
Following \citet{Emerick19} and \citet{Brauer25}, we calculate the total wind feedback energy by considering both the kinetic and thermal components of the ejected material as:
\begin{align}
    E_\mathrm{w} &= \frac{1}{2}M_\mathrm{w} v_\mathrm{w}^2 + E_\mathrm{th} \\
    &= \frac{1}{2}M_\mathrm{w} v_\mathrm{w}^2 + \frac{3}{2}\frac{M_\mathrm{w} k_\mathrm{B}T_\mathrm{eff}}{m_\mathrm{p}},
\end{align}
where $M_\mathrm{w}$ is the stellar wind mass loss, $v_\mathrm{w}$ is the wind velocity, $T_\mathrm{eff}$ is the effective temperature of the star, and $m_\mathrm{p}$ is the proton mass.
For our fiducial models, we set the wind velocity to $500\;\mathrm{km}\;\mathrm{s}^{-1}$ for massive stars ($M_\mathrm{ZAMS} \geq 8\;\mathrm{M}_\odot$) and $20\;\mathrm{km}\;\mathrm{s}^{-1}$ for lower-mass stars ($M_\mathrm{ZAMS} < 8\;\mathrm{M}_\odot$).
To investigate the impact of stronger feedback, we also explore models with a boosted wind velocity of $3000\;\mathrm{km\;s^{-1}}$ for massive stars.
The feedback energy is derived dynamically using stellar properties (mass loss and effective temperature) provided by \texttt{SEVN}.
For instance, assuming a typical effective temperature of $10^4\;\mathrm{K}$, a mass loss event of $0.01\;\mathrm{M_\odot}$ yields total feedback energies of approximately $6.5\times 10^{43}$, $2.5\times 10^{46}$, and $9.0\times 10^{47}\;\mathrm{erg}$ for wind velocities of $20$, $500$, and $3000\;\mathrm{km\;s^{-1}}$, respectively.
Our simulation suite with varying wind velocity $v_\mathrm{w}$ values (for massive stars) is summarized in Table \ref{tab:VMS_mass}.

Once the feedback energy is determined, we inject mass, thermal energy, and metals into the neighboring cells, following the method described in \citet{Simpson15}.
Specifically, we define a virtual $3^3\!$ cell-cloud centered on the star particle.
The feedback quantities are then distributed among the $4^3$ real cells that overlap with this virtual cloud, where the amount received by each cell is weighted by its volume overlap with the virtual cloud centered on the particle.

\subsubsection{Photoionization Heating (HII Region) Feedback} \label{subsubsec:HII}

We modify the method used in \citet{Goldbaum16} to implement the effect of photoionization heating by injecting thermal energy.
If a star is in the main sequence phase and its initial mass is larger than $5\;\mathrm{M}_\odot$, we calculate the Str{\"o}mgren radius:
\begin{equation}
    R_\mathrm{s} = \left( \frac{3Q}{4\pi\alpha_\mathrm{B}n^2} \right)^{1/3},
\end{equation}
where $Q$ is the ionizing photon rate, $\alpha_\mathrm{B} = 2.6\times10^{-13}\;\mathrm{cm}^3\;\mathrm{s}^{-1}$ is the case-B recombination coefficient at a fixed temperature of $10^4\;\mathrm{K}$, and $n$ is the hydrogen number density of gas in the cell containing a star particle. To determine $Q$, we use the same fitting formula and coefficients described in Eq. 2 and Table 1 in \citet{Fujii21}, which is obtained from OSTAR2002 \citep{Lanz03} with the assumption of solar metallicity and ignoring the stellar evolution.
The formula covers the mass range of $5\;\mathrm{M}_\odot \leq M_\mathrm{ZAMS} \leq 300\;\mathrm{M}_\odot$ and we use the value at $300\;\mathrm{M}_\odot$ if there is a star with $M_\mathrm{ZAMS} \geq 300\;\mathrm{M}_\odot$.

Following the method in \citet{Goldbaum16}, we compare the cell volume $V_\mathrm{c}$ containing the star particle and the volume of the Str{\"o}mgren sphere, $V_\mathrm{s} = (4/3)\pi R_\mathrm{s}^3$.
The cell containing the star particle is heated to a temperature of $T = 10^4\;\mathrm{K} \cdot f_\mathrm{V}$ if the cell temperature $T_\mathrm{c}$ is colder than $T$, where the volume filling factor $f_\mathrm{V}$ is defined as $f_\mathrm{V} = \texttt{min}[V_\mathrm{s}/V_\mathrm{c},\;1.0]$.
This approach is reasonable since the Str{\"o}mgren radius is considerably smaller than the cell size most of the time, as we simulate very dense gas cloud.
If there are multiple HII region feedback sources, each source contributes separately with the maximum temperature of $10^4\;\mathrm{K}$.
We expect that this HII region feedback can be improved in future works by employing an adaptive ray tracing method in \texttt{Enzo} \citep{Wise11} as adopted in \citet{Emerick19} and \citet{Brauer25}.

\subsubsection{Supernovae Feedback and Remnants Formation} \label{subsubsec:supernovae}

We implement supernova (SN) feedback for stars that reach the end of their lives with some specific conditions, as determined by the stellar evolution code \texttt{SEVN}.
Depending on the final mass of the star's CO or He core, \texttt{SEVN} triggers one of three supernova types:  (1) electron capture (ECSN), (2) core-collapse (CCSN), or (3) pair-instability (PISN).
The energy for each SN event is fixed at $10^{51}\;\mathrm{erg}$, which is injected as thermal energy regardless of the SN type.
This energy and metals are distributed among the neighboring cells using the same method as for stellar wind feedback, as described in Section \ref{subsubsec:winds}.
If the fallback fraction\footnote{The fallback fraction refers to the fraction of mass that gravitationally collapses onto the remnant behind the outgoing shock during the explosion.} is $1.0$, the CCSN event is not triggered and the star collapses directly into a black hole without SN feedback.

Following a SN event, the progenitor star particle is replaced by a compact remnant, according to the model adopted in \texttt{SEVN}.
After an ECSN, the star becomes a NS. 
White dwarfs (WDs) can form from lower-mass progenitors and may produce Type Ia supernovae via merger events. 
However, we do not consider these channels in this study, as our simulation timescale ($\lesssim 13\;\mathrm{Myr}$) is significantly shorter than the stellar lifetimes of such progenitors.
In contrast,  following a CCSN, a black hole (BH) or a neutron star (NS) can form; for these events, we adopt the delayed model of \citet{Fryer12}. 
The neutrino mass loss correction \citep{Lattimer89, Zevin20} is applied to a compact remnant, and the remnant receives a natal kick after SN events by the model suggested in \citet{Giacobbo20} (more information can be found in Appendix \ref{appendix:stellar-BH}).
Lastly, if a progenitor star meets the condition for a PISN, the explosion leaves no compact remnant.
PISNe occur when the final He-core mass ($M_\mathrm{He,f}$) is in the range of $64\leq M_\mathrm{He,f}/\mathrm{M}_\odot \leq 135$ \citep{Mapelli20, Spera17, Woosley17}.
VMSs with masses above this PISN mass gap do not explode and instead collapse directly into a BH.
This is the formation mechanism for IMBHs observed in our simulation suite.

\subsection{Intermediate-mass Black Hole (IMBH) Physics} \label{subsec:IMBH}

We define an IMBH as a BH with mass $M_\mathrm{BH} \geq 100\;\mathrm{M}_\odot$, and model their evolution and feedback using the massive black hole modules in \texttt{Enzo}, which implement gas accretion and feedback based on the algorithms of \citet{Kim11}.

\subsubsection{Gas Accretion} \label{subsubsec:accretion}

IMBHs in our simulations accrete gas from surrounding cells.
Their accretion rates, $\dot{M}_\mathrm{BH}$, are determined by the Eddington-limited Bondi-Hoyle-Lyttleton formula \citep{Hoyle39, Bondi52}:
\begin{align}
\dot{M}_\mathrm{BH} &= \texttt{min}(\dot{M}_\mathrm{B},\;\dot{M}_\mathrm{Edd}) \\
&= \texttt{min} \left( \frac{4\pi G^2M^2_\mathrm{BH}\rho_\mathrm{B}}{(v_\mathrm{rel}^2 + c_\mathrm{s}^2)^{3/2}},\;\frac{4\pi GM_\mathrm{BH}m_\mathrm{p}}{\epsilon_\mathrm{r}\sigma_\mathrm{T}c} \right),
\end{align}
where $v_\mathrm{rel}$ is the relative velocity between an IMBH and the gas cell, $c_\mathrm{s}$ is the sound speed of the cell containing an IMBH, $\epsilon_\mathrm{r} = 0.1$ is the radiative efficiency \citep{Shakura73}, and $\sigma_\mathrm{T}$ is the cross section of the Thomson scattering.
The Bondi radius is given by $R_\mathrm{B} = 2GM_\mathrm{BH}/c^2_\mathrm{s}$ and the density at the Bondi radius is estimated as
\begin{equation}
    \rho_\mathrm{B} = \rho_\mathrm{gas}\;\cdot\;\texttt{min}((\Delta x / R_\mathrm{B})^{1.5},\; 1.0), 
\end{equation}
where $\Delta x$ is the size of the cell the IMBH resides in \citep{Wang10, Kim11, Kim19, Lee23}.
Our most refined cell size of $0.04\;\mathrm{pc}$ is sufficient to resolve the Bondi radius of a $1000\;\mathrm{M}_\odot$ IMBH, which is typically $\sim 0.1\;\mathrm{pc}$ in our simulations.
To satisfy local mass conservation, the gas mass equals to the accreted mass into a BH is subtracted from the sphere of gas cells with the radius of $0.16\;\mathrm{pc}$ (corresponding to the four smallest cell width).

\subsubsection{Tidal Disruption Accretion} \label{subsubsec:TDA}

While \citet{Lee23} modeled TDE accretion indirectly using analytical estimates of the stellar density and velocity dispersion around the massive black hole (MBH), our direct $N$-body approach allows us to explicitly track individual stellar orbits.
We assume that a TDE occurs when the separation between an IMBH and a star is smaller than the tidal disruption radius \citep{Kochanek92, Rantala23},
\begin{equation} \label{eq_r_TDE}
    R_\mathrm{TDE} = 1.3\; \left( \frac{M_\star + M_\mathrm{BH}}{M_\star} \right)^{1/3}r_\star.
\end{equation}
Upon disruption, we assume the IMBH instantaneously accretes half of the stellar mass, consistent with previous numerical studies \citep[e.g.,][]{Rizzuto23, Rantala23, Lee25}.
The remaining half is ejected as gas and distributed uniformly into the $3^3\!$ cells surrounding the IMBH, where it contributes to the local gas reservoir for potential star formation or future accretion.

\subsubsection{Thermal Feedback} \label{subsubsec:thermal feedback}

AGNs with high accretion rates comparable to the Eddington limit are thought to operate in the quasar mode, where feedback is primarily driven by radiation pressure \citep{Dubois12, Cielo18}.
In our simulation suites, IMBHs accrete gas at the Eddington rate for the majority of their evolution. 
This high accretion efficiency arises from our dense gas initial conditions and the initially weak stellar feedback, as most stars remain on the main sequence during the rapid IMBH formation via runaway collisions.
Consequently, we employ a thermal feedback prescription to represent the quasar mode for these IMBHs, adapting a method traditionally developed for SMBHs.

We calculate the thermal energy ejection rate via AGN feedback as follows:
\begin{equation}
    \dot{E}_\mathrm{AGN} = \epsilon_\mathrm{ec}\epsilon_\mathrm{r}\dot{M}_\mathrm{BH} c^2,
\end{equation}
where $\epsilon_\mathrm{ec} = 10^{-4}$ is our  fiducial energy coupling constant.\footnote{In most simulations describing SMBH thermal feedback ($M_\mathrm{BH} > 10^6\;\mathrm{M}_\odot$), $\epsilon_\mathrm{ec}$ ranges from 0.05 \citep{Springel05, DiMatteo05} to 0.15 \citep{Booth09, Dubois12}.  We adopted a lower value since our simulation depicts IMBHs ($M_\mathrm{BH} < 10^4\;\mathrm{M}_\odot$).
For a $5000\;\mathrm{M_\odot}$ IMBH accreting at the Eddington limit, the resulting energy ejection rate is $6.3\times10^{37}\;\mathrm{erg\;s^{-1}}$, which is small enough to ensure that the IMBH does not unphysically quench local star formation or evacuate the cluster core prematurely.}
The calculated thermal feedback energy is injected into the surrounding spherical region with the radius of $0.1\; \mathrm{pc}$.
While the adopted thermal feedback prescription for IMBHs is exploratory and has only a marginal impact on IMBH growth in the present study, we will examine the effects of varying feedback parameters in future work.

\begin{figure*}[t]
    \vspace{-1mm}
    \centering
    \includegraphics[width=\textwidth]{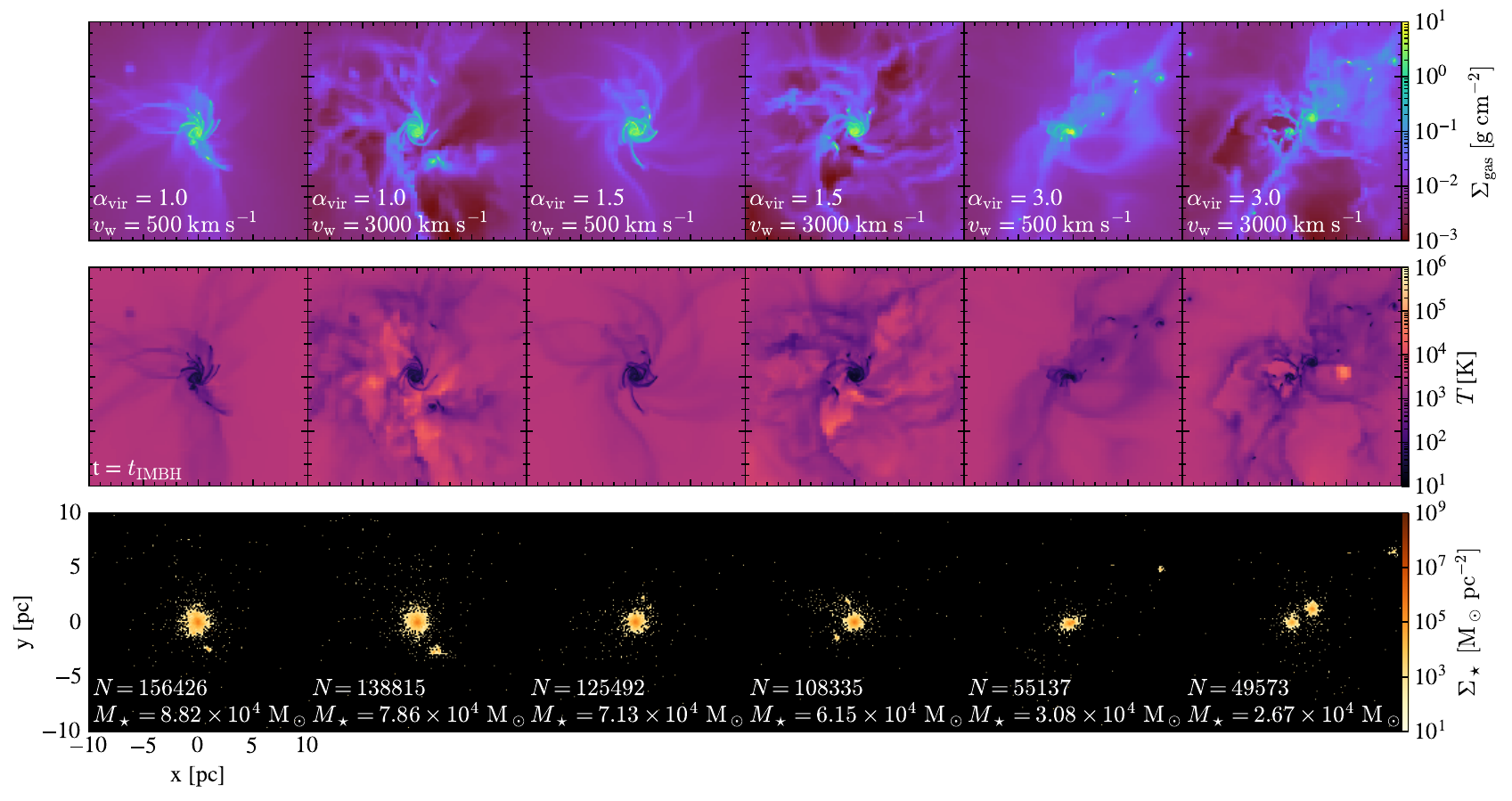}
    \caption{
    Snapshots of the central $20\;\mathrm{pc}$ region from the simulation suite, each taken immediately before a VMS collapses into an IMBH.
    Each row shows gas surface density (\textit{top}), density-weighted gas temperature (\textit{middle}), and stellar surface density (\textit{bottom}).
    The total number of stars and stellar mass within the simulation domain are indicated.
    For a fixed virial parameter, models with stronger stellar wind feedback exhibit hot, low-density cavities, while the ones with weaker feedback form more massive clusters.
    See Section \ref{subsec:formation of massive star clusters} for more information.
    }
    \label{density_temp_particle}
    \vspace{2mm}
\end{figure*}

\subsection{Mergers Amongst Stars and Compact Objects} \label{subsec:merger}

Since we are dealing with direct $N$-body method without softening between particles, there is no resolution limit in gravity.
In our simulations, particles are not treated as point masses; instead, each particle represents an object with a finite radius.
We classify three distinct types of mergers: (1) stellar mergers (star-star), (2) star-compact object mergers, and (3) compact object-compact object mergers.
Close encounters are integrated using the \texttt{SDAR} library and we trigger a merger based on two distinct conditions, which are checked at every \texttt{SDAR} timestep ($\Delta t_\mathrm{SDAR}$):
\begin{itemize}
    \item If the current separation between particles ($r_\mathrm{sep}$) is already within the merger radius ($r_\mathrm{sep} \leq r_\mathrm{merger}$).
    \item If the particles are currently separated by more than the merger radius ($r_\mathrm{sep} > r_\mathrm{merger}$), but they will pass within it ($r_\mathrm{peri} \leq r_\mathrm{merger}$) before the next timestep is complete ($t_\mathrm{peri} \leq \Delta t_\mathrm{SDAR}$), where $r_\mathrm{peri}$ is the periapsis distance and $t_\mathrm{peri}$ is the time to reach periapsis.
\end{itemize}
Furthermore, to account for general relativistic effects, we incorporate the orbit averaged post-Newtonian corrections for bound few-body systems\footnote{For few-body systems composed of more than two particles, we identify all bound particle pairs within the hierarchical binary tree of \texttt{SDAR} and apply the post-Newtonian corrections to their relative motions. This approach can be less accurate in strongly interacting three-body systems compared to full post-Newtonian equations of motion.}.
These corrections can harden the system and accelerate potential mergers (more information can be found in Appendix \ref{appendix:PN}).

We determine stellar radii using \texttt{SEVN} for stars in the mass range of $2.2\;\mathrm{M}_\odot \leq M_\star \leq 600\;\mathrm{M}_\odot$.
For stars outside this range, we calculate the radius by assuming a constant stellar density.
For low-mass stars with $M_\star \leq 2.2\;\mathrm{M}_\odot$, we adopt the solar density, yielding $r_\star = \mathrm{R}_\odot\cdot(M_\star/\mathrm{M}_\odot)^{1/3}$.
For VMSs with $M_\star \geq 600\;\mathrm{M}_\odot$, we extrapolate using the density of a $600\;\mathrm{M}_\odot$ star, resulting in $r_\star = \mathrm{R}_{600}\cdot(M_\star/600\;\mathrm{M}_\odot)^{1/3}$, where $\mathrm{R}_{600}$ is a radius of a $600\;\mathrm{M}_\odot$ star.
When stellar mergers take place, we assume that chemical mixing of two stars occur immediately without mass loss and stellar properties of the remnant are determined by the merger process in \texttt{SEVN}.
The merger remnant is assigned the evolutionary stage and fractional lifetime of the more evolved progenitor.
Further details on how \texttt{SEVN} treats stellar mergers can be found in \citet{Iorio23}.

In contrast, to describe interactions between compact objects (WDs, NSs, and stellar-mass BHs) and stars, we extend the treatment of TDEs described in Section \ref{subsubsec:TDA}.  
Specifically, any merger between a star and these stellar-mass remnant is treated as a TDE.\footnote{In this case, $M_\mathrm{BH}$ in Eq. \ref{eq_r_TDE} is replaced by the mass of the remnant, $M_\mathrm{rem}$.}
Finally, a merger between two compact objects is triggered only if their separation falls below the innermost stable circular orbit (ISCO), which we approximate as $R_\mathrm{isco} = 6GM_\mathrm{rem}/c^2$ using the Schwarzschild metric.
However, we note that while IMBH-driven TDEs are frequent, no interactions involving these stellar-mass remnants occurred during our simulations.
Although a small population ($\gtrsim100$) of stellar-mass BHs formed across our runs, the simulation runtime of $\sim13\;\mathrm{Myr}$ was too short for these rare dynamical channels (stellar-mass remnant TDEs or compact object mergers) to manifest (see Section \ref{subsec:after VMS}).

The position and velocity of a merger remnant are determined by the center-of-mass of the two particles before the merger.
Furthermore, if both objects are BHs, the mass, spin, and recoil kick of the merger product can be modeled to account for gravitational wave emission.
However, we reserve this detailed treatment for future work, as no such events occurred in our current simulations.

\begin{figure}[t]
    \centering
    \includegraphics[width=\columnwidth]{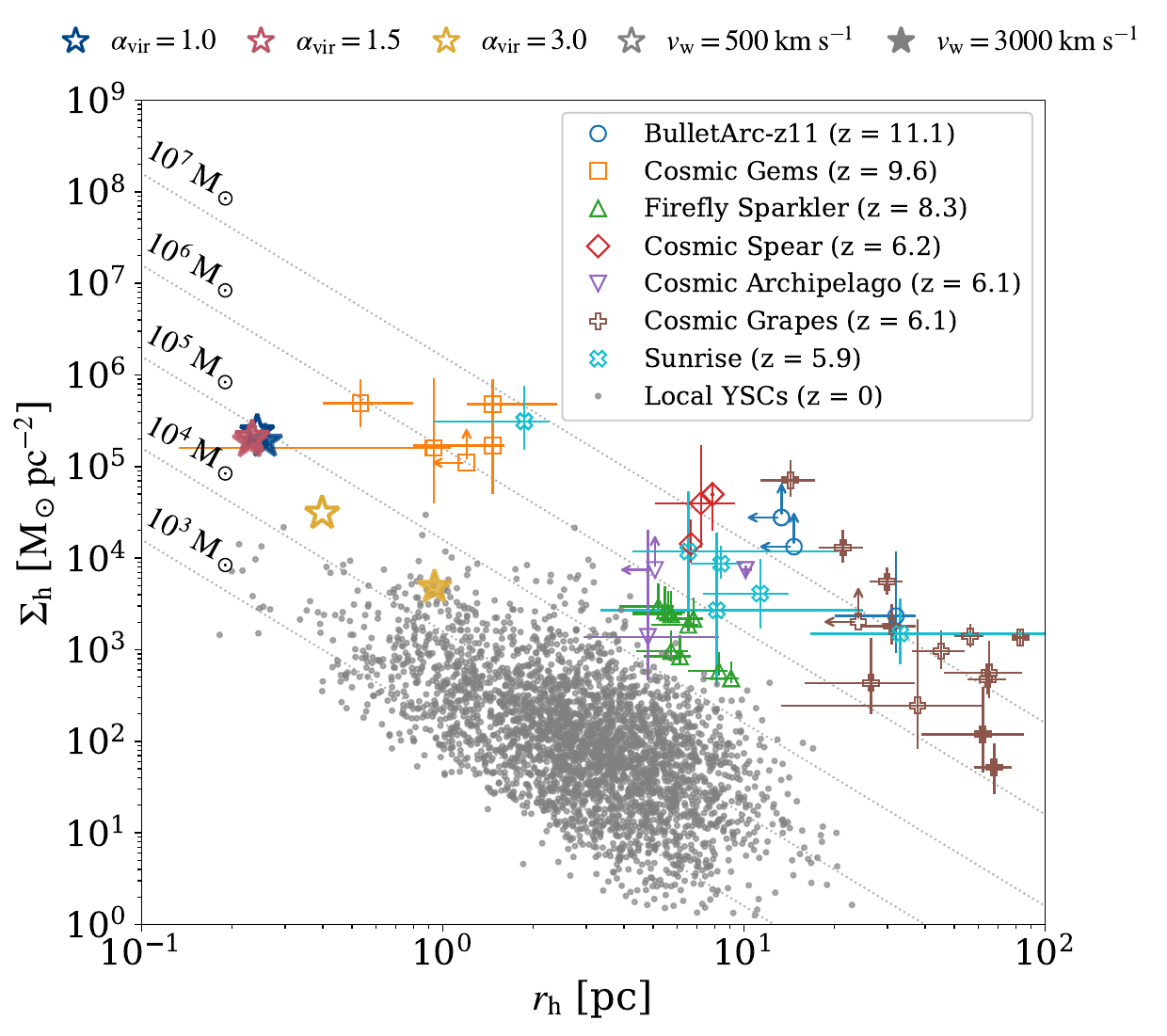}
    \caption{
    Half-mass radius versus stellar surface density for the simulated star clusters, compared with observational samples from high-redshift galaxies and the local Universe.
    Simulated clusters are represented by star-shaped markers: different colors indicate different initial virial parameters ($\alpha_\mathrm{vir} = 1.0$ in blue, 2.0 in red, and 3.0 in gold), while open and filled markers denote models with fiducial ($v_\mathrm{w} = 500\;\mathrm{km\;s^{-1}}$) and strong ($v_\mathrm{w} = 3000\;\mathrm{km\;s^{-1}}$) stellar wind feedback, respectively.
    Observational data include BulletArc-z11 \citep{Bradac25}, Cosmic Gems \citep{Adamo24, Messa25b}, Firefly Sparkler \citep{Mowla24}, Cosmic Spear \citep{Abdurrouf25}, Cosmic Archipelago \citep{Messa25a}, Sunrise \citep{Vanzella23}, and local young star clusters \citep{Brown21}. The simulated clusters are compact ($r_\mathrm{h} \lesssim 1\;\mathrm{pc}$) and exhibit high surface densities ($\sim10^5\;\mathrm{M_\odot\;pc^{-2}}$), comparable to the properties of the Cosmic Gems clusters.
    Note that the two simulated clusters with $\alpha_\mathrm{vir} = 3.0$ show lower surface density than the others due to their high initial turbulence.
    See Section \ref{subsec:formation of massive star clusters} for more information.
    }
    \label{surface density}
    \vspace{3mm}
\end{figure}

\hspace{1mm}

\section{Results} \label{sec:results}

We perform a suite of simulations with varying wind feedback strengths and initial virial parameters, evolving each until a VMS collapses into an IMBH (see Table \ref{tab:VMS_mass}).
Given the short lifetime of VMSs, these simulations cover a duration of $<4\;\mathrm{Myr}$.
Subsequently, we select two representative simulations and extend them to $\sim 13\;\mathrm{Myr}$ to investigate the co-evolution of the star cluster and the IMBH.
Unless otherwise noted, all time references denote the time elapsed since the first star formation event.
In this work, sub-clusters form from the fragmenting gas and subsequently collapse to form a single massive star cluster in every model. 
We identify these sub-clusters morphologically rather than employing a formal clustering algorithm; they are clearly distinguishable as dense stellar groupings in the snapshots (e.g., the two sub-clusters visible in the sixth column of the bottom panel of Figure \ref{density_temp_particle}).

\subsection{Formation of Massive Star Clusters} \label{subsec:formation of massive star clusters}

Figure \ref{density_temp_particle} presents the gas surface density, density-weighted gas temperature, and stellar surface density at the epoch immediately preceding the collapse of the VMS in each model.
As the initial gas density profile is sufficiently concentrated to support star formation in the core, a central star cluster forms in every model, with the exception of the case with $v_\mathrm{w} = 3000\;\mathrm{km\;s^{-1}}$ and $\alpha_\mathrm{vir} = 3.0$, in which two central clusters are in the process of merging.
However, variations in initial virial parameter and wind feedback strength lead to considerable differences in the gas distribution and the timing of the VMS collapse.
In the gas density projections of models with strong stellar wind feedback (e.g., $v_\mathrm{w} = 3000\;\mathrm{km\;s^{-1}}$), we observe localized low-density regions, particularly in the off-center areas, caused by escaping massive stars.
In the core of dense clusters, frequent close encounters and three-body interactions can eject massive stars at high velocities \citep{Fujii11, Fujii22}.
As these high-velocity `escapers' traverse the simulation volume, their stellar winds clear out the surrounding gas.
Since we implement stellar wind feedback as thermal energy injection, these feedback-induced low-density cavities also correspond to high-temperature regions in the temperature projections.

\begin{figure}[t]
    \centering
    \includegraphics[width=\columnwidth]{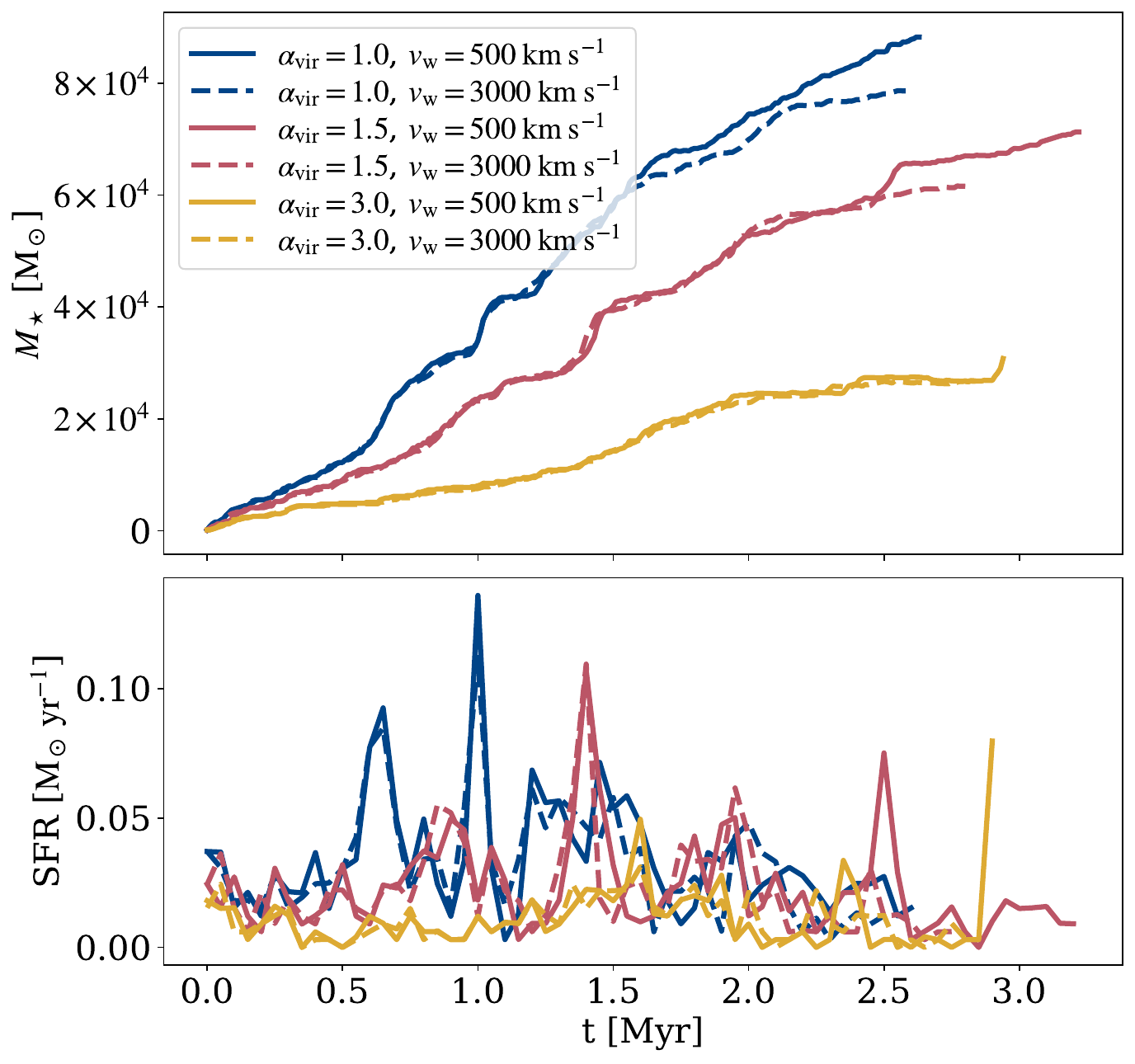}
    \caption{
    Evolution of the total stellar mass within the simulation domain (\textit{top}) and star formation rates (\textit{bottom}).
    The less turbulent models (with lower virial parameter $\alpha_\mathrm{vir}$) exhibit bursty star formation at earlier epochs.
    In contrast, variations in the wind feedback strength ($v_\mathrm{w}$) do not produce significant differences in either the total stellar mass or the star formation history when the same $\alpha_\mathrm{vir}$ is used.  
    See Section \ref{subsec:formation of massive star clusters} for more information.
    }
    \label{combined_mass_SFR}
\end{figure}

To investigate the structural properties of simulated clusters and compare them with observations, in Figure \ref{surface density} we present their half-mass radii and stellar surface densities\footnote{Stellar surface density is defined as $\Sigma_\mathrm{h} = {M_\mathrm{cl}}/({2 \pi r_\mathrm{h}^2})$, where $M_\mathrm{cl}$ is the cluster mass and $r_\mathrm{h}$ is the half-mass radius.} at the moment of IMBH formation, along with the observed clusters from both high-redshift galaxies and the local Universe.
The properties of the simulated clusters are measured immediately prior to the collapse of the VMS.
The observational comparison samples include the BulletArc-z11 \citep{Bradac25}, Cosmic Gems \citep{Adamo24, Messa25b}, Firefly Sparkler \citep{Mowla24}, Cosmic Spear \citep{Abdurrouf25}, Cosmic Archipelago \citep{Messa25a}, Cosmic Grapes \citep{Fujimoto25}, and the Sunrise arc \citep{Vanzella23}, as well as local young star clusters \citep[YSCs;][]{Brown21}.
For the local YSCs, only clusters with estimated ages younger than $100\;\mathrm{Myr}$ are selected.

\begin{figure*}[t]
    \centering
    \includegraphics[width=\textwidth]{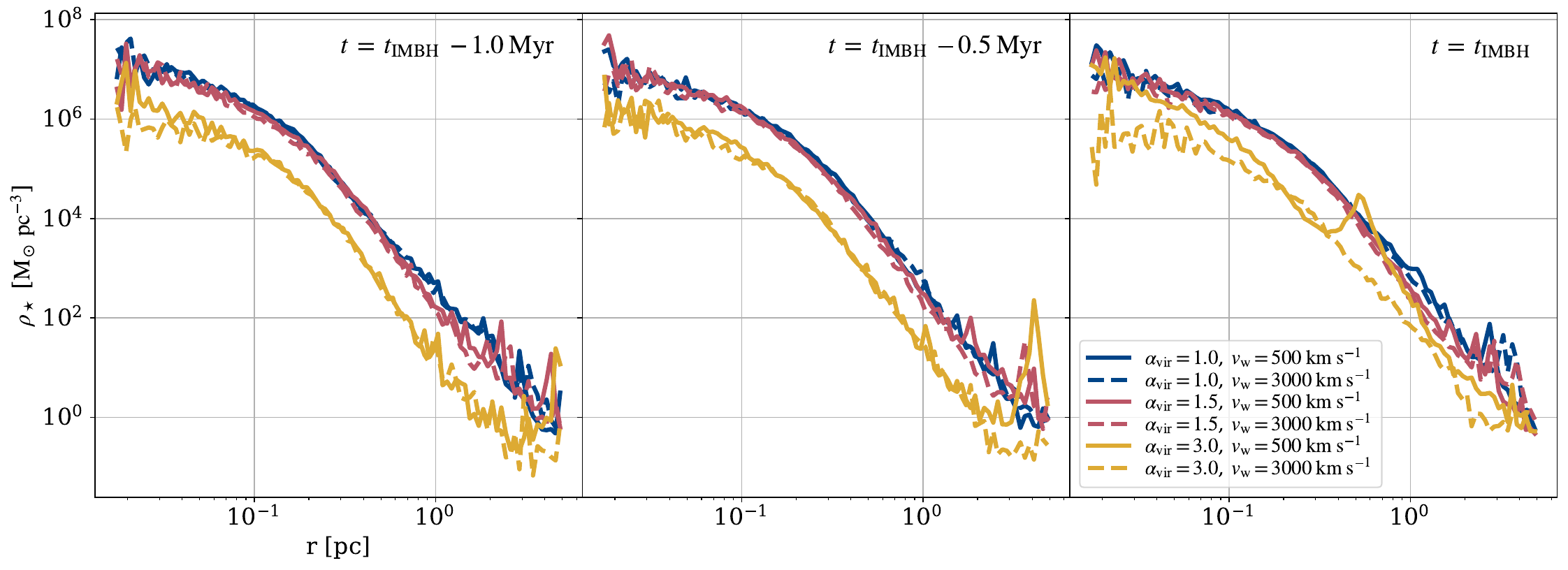}
    \caption{
    Stellar density profiles measured from the density center at three epochs: 1.0, 0.5, 0.0 Myr before the formation of an IMBH.
    Here, the profile center is defined as the location of the maximum density peak within $0.1\;\mathrm{pc}$ of the VMS (which is about to collapse into an IMBH).
    For models with $\alpha_\mathrm{vir} = 1.0$ and $1.5$, densities in the central $r < 0.1\;\mathrm{pc}$ region reach $\sim10^6\;\mathrm{M_\odot\;pc^{-3}}$ --- the threshold at which runaway collisions can occur \citep{Ardi08, Fujii24} ---  at all epochs shown.
    In contrast, the most initially turbulent models ($\alpha_\mathrm{vir} = 3.0$) reach this threshold only in the innermost regions and exhibit lower densities overall, leading to the formation of significantly less massive VMSs ($M_\mathrm{VMS} < 600\;\mathrm{M_\odot}$; see also Table \ref{tab:VMS_mass}).
    See Section \ref{subsec:VMS evolution} for more information.
    }
    \label{stellar_density_profile}
    \vspace{2mm}
\end{figure*}

Most local YSCs have masses below $10^4\;\mathrm{M_\odot}$ and surface densities of $\sim10^3\;\mathrm{M_\odot\;pc^{-2}}$.
In contrast, due to their compact, sub-parsec sizes, our simulated clusters generally exhibit higher densities than the majority of observed systems --- with two possible exceptions of $\alpha_\mathrm{vir} = 3.0$ (the most turbulent initial conditions, with lower surface densities of $\sim10^4\;\mathrm{M_\odot\;pc^{-2}}$).
Our simulated clusters are more massive ($\sim10^5\;\mathrm{M_\odot}$), and more comparable to high-redshift observational samples.
Among observed systems, the Cosmic Gems clusters \citep[spectroscopically confirmed at $z = 9.6$;][] {Messa25b} represents a striking counterpart to our fiducial models, displaying high stellar surface densities of $10^5-10^6\;\mathrm{M_\odot\;pc^{-2}}$.
Beyond density, the Cosmic Gems clusters share other key similarities with our simulations:
they are young ($7 - 30\;\mathrm{Myr}$), metal-poor ($Z < 0.1\;Z_\odot$), and reside in a mini-quenched state \citep{Messa25b}.
Furthermore, some clusters from the Cosmic Spear, Cosmic Grapes, and Sunrise arc also exhibit high surface densities similar to our models.
We will later argue that such dense star clusters at high redshift are promising environments for IMBH formation via runaway collisions, as are also suggested by recent studies \citep[e.g.,][]{Mayer25, Rantala25a, Lahen25_merger, vanDonkelaar26}.

Figure \ref{combined_mass_SFR} shows the total stellar mass growth and the star formation rate in each simulation.
We observe distinct differences in stellar mass growth across simulations with different initial virial parameters, $\alpha_\mathrm{vir}$.
The total stellar mass in models with a lower virial parameter  is significantly larger than in those with a higher virial parameter.
This is expected, as less turbulent models undergo more efficient gravitational collapse.
The trend is also evident in the star formation rate;
The less turbulent models exhibit intense bursts of star formation at early epochs, driven by rapid gravitational collapse.

In contrast, variations in the wind feedback strength parameter, $v_\mathrm{w}$, do not produce significant differences when the same $\alpha_\mathrm{vir}$ is used. 
Minor divergence begins to appear at later times ($t > 1.5\;\mathrm{Myr}$), particularly in the models with $\alpha_\mathrm{vir} = 1.0$ and $1.5$. 
During the early phase ($t < 1.5\;\mathrm{Myr}$), stars remain on the main sequence, and mass loss is negligible according to the stellar evolution model  in \texttt{SEVN}.
As the system evolves, wind feedback begins to impact the surrounding gas, leading to suppressed star formation in models with larger $v_\mathrm{w}$.
However, the overall suppression is marginal because the stellar wind feedback is insufficient to disrupt the gravitational potential of massive molecular clouds ($M_\mathrm{cloud} \gtrsim 10^5\;\mathrm{M_\odot}$), a finding consistent with \citet{Polak24}.
In the $\alpha_\mathrm{vir} = 3.0$ cases, differences between wind feedback strengths are negligible, as the number of feedback sources (i.e., massive stars) is relatively smaller than in the other models.

\subsection{Formation and Evolution of VMSs} \label{subsec:VMS evolution}

Successive stellar mergers in the dense cores of star clusters can lead to the formation of VMSs \citep{Portegies02, Portegies04}. 
According to the stellar evolution model we adopted in \texttt{SEVN}, a VMS directly collapses into an IMBH if its final helium core mass exceeds $135\;\mathrm{M_\odot}$ (see Section \ref{subsubsec:supernovae}).
We find that the central massive clusters in all our simulations are sufficiently dense to trigger this pathway, producing VMSs massive enough to undergo direct collapse into IMBHs.
Here, we investigate the formation of the VMS in each simulation and track its evolution until this collapse occurs.

Figure \ref{stellar_density_profile} illustrates the evolution of stellar density profiles.
Three epochs are selected (1.0, 0.5, 0.0 Myr before an IMBH forms) to capture the final stages of the VMS lifetime, when stellar mergers occur most frequently.
As shown in previous studies \citep{Ardi08, Fujii24}, a core density exceeding $\sim10^6\;\mathrm{M_\odot\;pc^{-3}}$ is a prerequisite for triggering runaway stellar collisions.
Our results confirm that all simulated clusters satisfy this criterion across the plotted epochs, indicating that VMS formation is a natural outcome in these environments.
Specifically, the lower-turbulence models ($\alpha_\mathrm{vir} = 1.0$ and 1.5) sustain extremely high densities above the $10^6\;\mathrm{M_\odot\;pc^{-3}}$ threshold throughout the inner $0.1\;\mathrm{pc}$, and peak at $\sim10^7\;\mathrm{M_\odot\;pc^{-3}}$.
In contrast, the high-turbulence models ($\alpha_\mathrm{vir} = 3.0$) show significantly lower densities.  
They attain the critical density only in the immediate vicinity of the center.
Consequently, the denser environments in the low-turbulence models are far more conducive to the rapid sequence of mergers required to assemble massive VMSs.

\begin{figure}[t]
    \centering
    \includegraphics[width=\columnwidth]{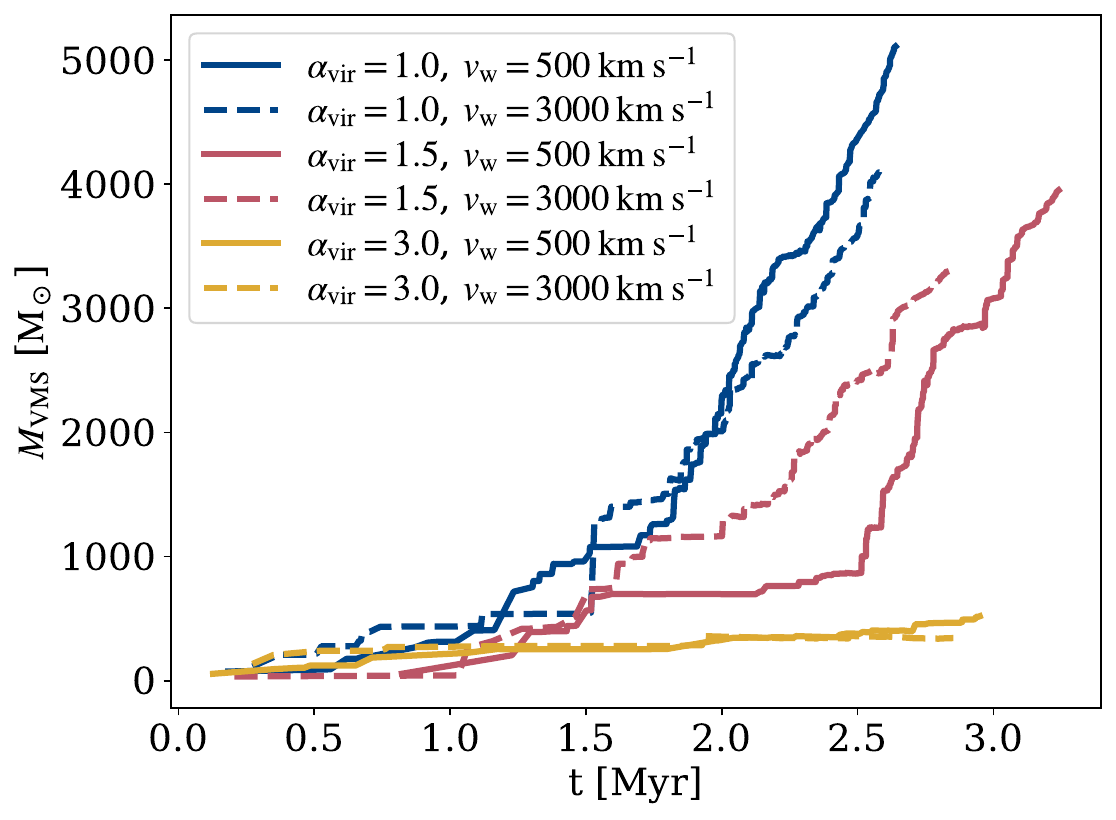}
    \caption{
    Mass evolution of the VMS that later becomes an IMBH in each model. VMSs exceeding $3000\;\mathrm{M_\odot}$ form in all cases except in the most turbulent models ($\alpha_\mathrm{vir} = 3.0$), where initial high turbulence hinders bursty star formation (Figure \ref{combined_mass_SFR}). Additionally, models with stronger wind feedback consistently yield less massive VMSs.
    See Section \ref{subsec:VMS evolution} for more information.
    }
    \label{VMS_evolution}
\end{figure}

The mass evolution of the VMS that later collapses into an IMBH in each model is depicted in Figure \ref{VMS_evolution}.
As discussed earlier, the rapid growth is driven by successive stellar collisions in every simulation.
For models with $\alpha_\mathrm{vir} = 1.0$ and $1.5$, the collision process exhibits true runaway characteristics, where the mass growth accelerates over time as successive mergers occur.
Conversely, for the models with $\alpha_\mathrm{vir} = 3.0$, the mass evolution is noticeably milder and more linear due to the highly turbulent initial gas cloud and the resulting lower star formation rate.

The final VMS masses, measured immediately before collapsing into an IMBH, are listed in Table \ref{tab:VMS_mass}.
These VMSs tend to be one of the most massive ones in the simulation volume, and at the end of the simulation only one IMBH is born in each model.  
Unlike the other runs where $M_\mathrm{VMS}$ exceeds $3000\;\mathrm{M_\odot}$, the most turbulent models ($\alpha_\mathrm{vir} = 3.0$) produce significantly less massive VMSs ($M_\mathrm{VMS} < 600\;\mathrm{M_\odot}$), a consequence of their lower core densities observed in Figure \ref{stellar_density_profile}.
We also notice that, at fixed initial virial parameter, stronger wind feedback consistently leads to slightly lower final VMS masses.
In our simulations, strong wind feedback from massive stars heats and expels surrounding gas, suppressing both the overall star formation rate and the final VMS mass. 
However, it does not  entirely inhibit the formation of the VMS.
We attribute this resilience to the high initial density of the gas profile: while stellar wind feedback affects the system, it is insufficient to disrupt the central dense core where the majority of stars form and runaway collisions occur.

\begin{figure}[t]
    \centering
    \includegraphics[width=\columnwidth]{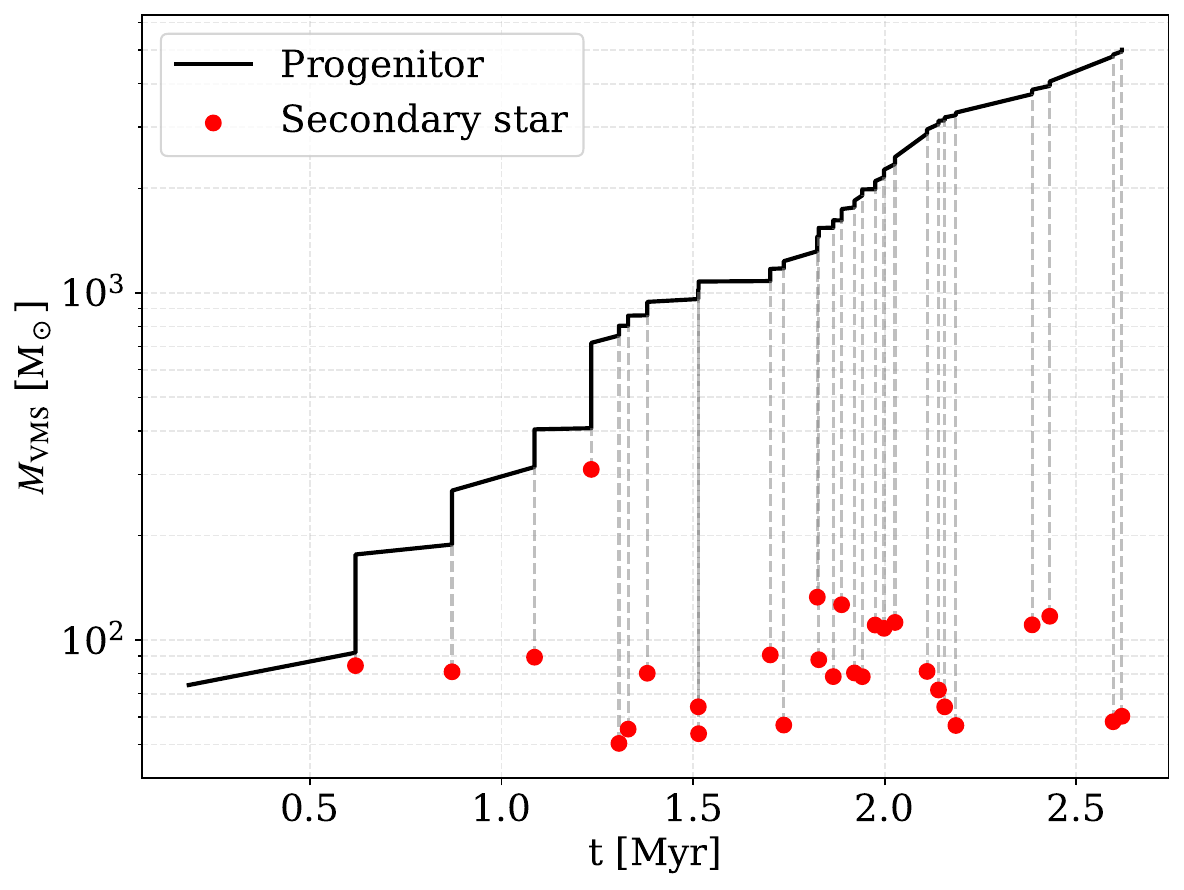}
    \caption{
    Merger tree of the VMS that later becomes an IMBH in the model with $\alpha_\mathrm{vir} = 1.0$ and $v_\mathrm{w} = 500\;\mathrm{km\;s^{-1}}$. Of the 2543 total stellar mergers, only the 28 events involving merging stars more massive than $50\;\mathrm{M_\odot}$ are shown. The VMS progenitor, with an initial mass of $74.2\;\mathrm{M_\odot}$, grows to $5108\;\mathrm{M_\odot}$ through successive stellar mergers.
    See Section \ref{subsec:VMS evolution} for more information.
    }
    \label{merger_tree}
\end{figure}

In Figure \ref{merger_tree}, the merger tree of the VMS that later becomes an IMBH in the run with $\alpha_\mathrm{vir} = 1.0$ and $v_\mathrm{w} = 500\;\mathrm{km\;s^{-1}}$ is presented.
Note that the figure displays only the 28 merger events where the merging star is more massive than $50\;\mathrm{M_\odot}$.
Over a total lifetime of $2.5\;\mathrm{Myr}$, 2543 stellar mergers drive the rapid growth of the VMS from an initial mass of $74.2\;\mathrm{M_\odot}$ to a final mass of $5108\;\mathrm{M_\odot}$.
The VMS exits the main sequence at an age of $\sim2.2\;\mathrm{Myr}$, at which point its radius begins to increase significantly, reaching a maximum of $\sim 30\;\mathrm{AU}$ at the end of its life.
However, we find that nearly half of all merger events occur during the main sequence phase (see Table \ref{tab:VMS_mass}). 
This indicates that the extreme central density of the star cluster is the primary driver of runaway collisions, rather than the increased cross-section of the VMS during its post-main sequence expansion.
During the early evolutionary stage of VMS ($t<1.2\;\mathrm{Myr}$), only 14 mergers occur, bringing the VMS mass to $407\;\mathrm{M_\odot}$;
at this point, the central massive star cluster has not yet fully formed.
At $t \sim 1.2\;\mathrm{Myr}$, two sub-clusters merge to form a single, massive star cluster.
During this event, the two massive VMSs ($407\;\mathrm{M_\odot}$ and $310\;\mathrm{M_\odot}$) merge.
Following the assembly of the central cluster, a burst of star formation occurs (as shown in Figure \ref{combined_mass_SFR}), triggering intense runaway stellar collision onto the central VMS.

\subsection{After the VMSs Collapse: Formation and Evolution of IMBHs} \label{subsec:after VMS}

\begin{figure}[t]
    \centering
    \includegraphics[width=\columnwidth]{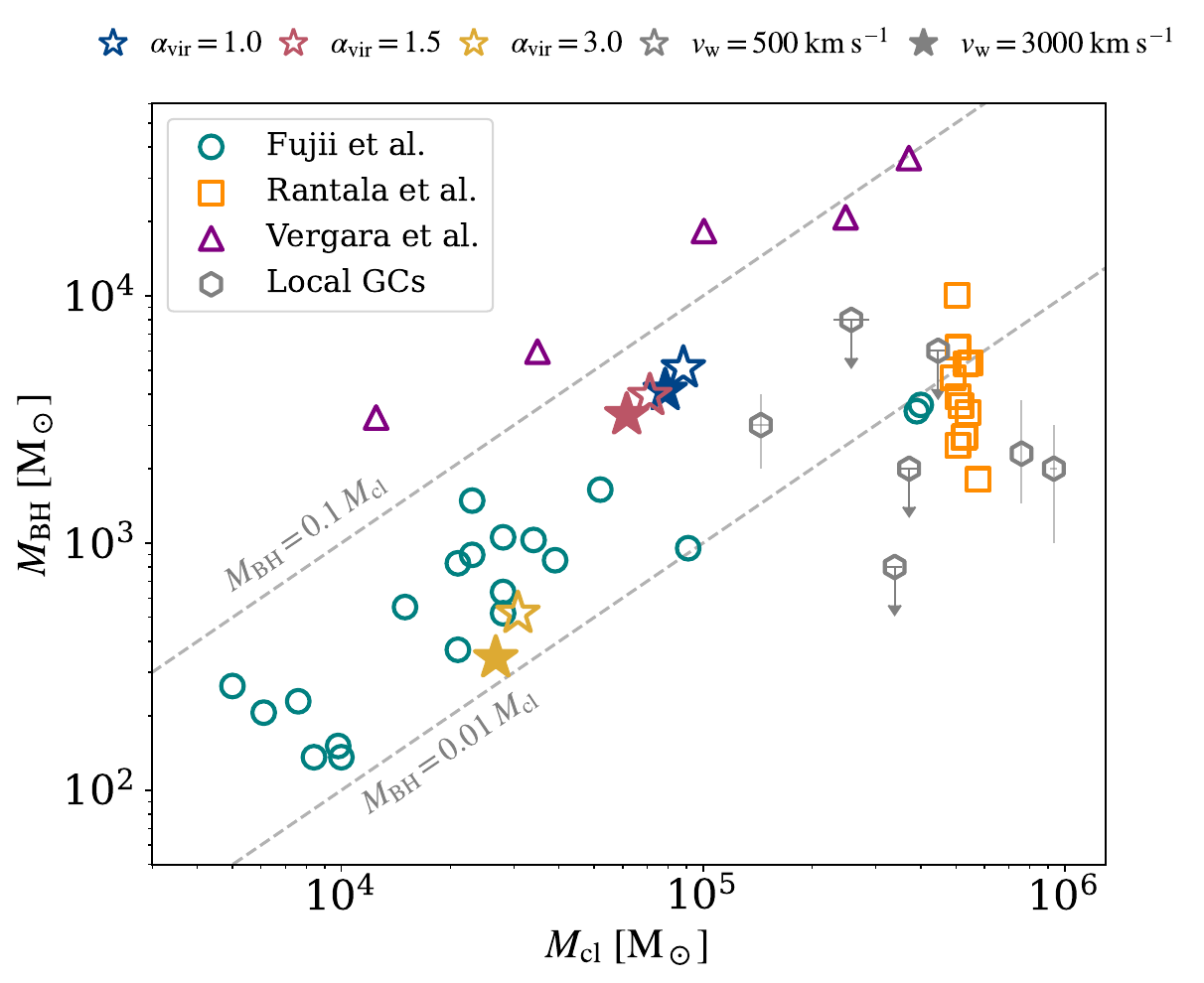}
    \caption{
    Relationship between the IMBH mass ($M_\mathrm{BH}$) and host cluster mass ($M_\mathrm{cl}$).
    Our simulation results are shown immediately following the collapse of the VMS into an IMBH, with the cluster mass defined as the total stellar mass within the simulation volume. 
    The same marker styles as in Figure \ref{surface density} are used to represent the simulated clusters in this work.
    For comparison, we include simulation data from \citet{Fujii24, Rantala25b, Vergara25} and observational estimates for local globular clusters \citep{Lutzgendorf13, Kiziltan17}. 
    While the IMBH masses from this work and \citet{Fujii24} typically fall within 1\% -- 10\% of the host cluster mass, the extremely dense models of \citet{Vergara25} yield even higher mass ratios.
    Conversely, both the hierarchical merger models of \citet{Rantala25b} and local GCs generally exhibit mass fractions below 1\%.
    See section \ref{subsec:after VMS} for more information.
    }
    \label{fig:mcl_mbh}
\end{figure}

In this section, we focus on the evolution of the IMBH following the collapse of the merger-induced VMS in two selected models.
In every model presented in the previous section (Section \ref{subsec:VMS evolution}), the VMS is sufficiently massive to avoid the PISN mass gap, resulting in direct collapse into an IMBH \citep{Heger02, Spera17}.
The relationship between the host cluster mass and the resulting IMBH mass is shown in Figure \ref{fig:mcl_mbh}.
For comparison, we include data from other gas-cloud-to-cluster simulations \citep{Fujii24} and isolated star cluster simulations \citep{Rantala25b, Vergara25}.
We also plot observational estimates for Milky Way globular clusters (GCs) estimated to harbor IMBHs, including NGC 1851, NGC 1904 (M 79), NGC 5694, NGC 5824, NGC 6093 (M 80), and NGC 6266 (M 62) \citep{Lutzgendorf13}, as well as NGC 104 (47 Tuc) \citep{Kiziltan17}.

Most simulated IMBHs from \citet{Fujii24} and this work fall within the range of 1\% -- 10 \% of the total cluster mass.
Models by \citet{Vergara25}, which employ extremely dense initial conditions (central densities up to $\sim10^{10}\;\mathrm{M_\odot}\;\mathrm{pc^{-3}}$), produce IMBHs exceeding 10\% of the cluster mass.
Conversely, the results from \citet{Rantala25b} show lower mass ratios;
in their hierarchical merging scenario, runaway collisions occur in individual sub-clusters, and the resulting IMBHs often fail to merge or are ejected via gravitational recoil.
Local GCs generally exhibit generally lower $M_\mathrm{BH}/M_\mathrm{cl}$ ratios than those found in simulations.
It is important to note that because GCs are ancient systems that have lost significant stellar mass through tidal stripping and evaporation \citep{Baumgardt03, Gieles11}, their mass ratios at birth were likely even lower than observed today, highlighting a persistent discrepancy between runaway collision simulations and local observations.

We investigate the subsequent evolution of these IMBHs until the continuous supernova feedback expels most of the remaining gas from the simulation domain.
During the VMS phase, we consider stellar mergers as the sole mechanism for mass growth.
However, following the collapse into an IMBH, it can grow via Eddington-limited gas accretion (Section \ref{subsubsec:accretion}) and tidal disruption accretion (Section \ref{subsubsec:TDA}).

\begin{figure}[t]
    \centering
    \includegraphics[width=\columnwidth]{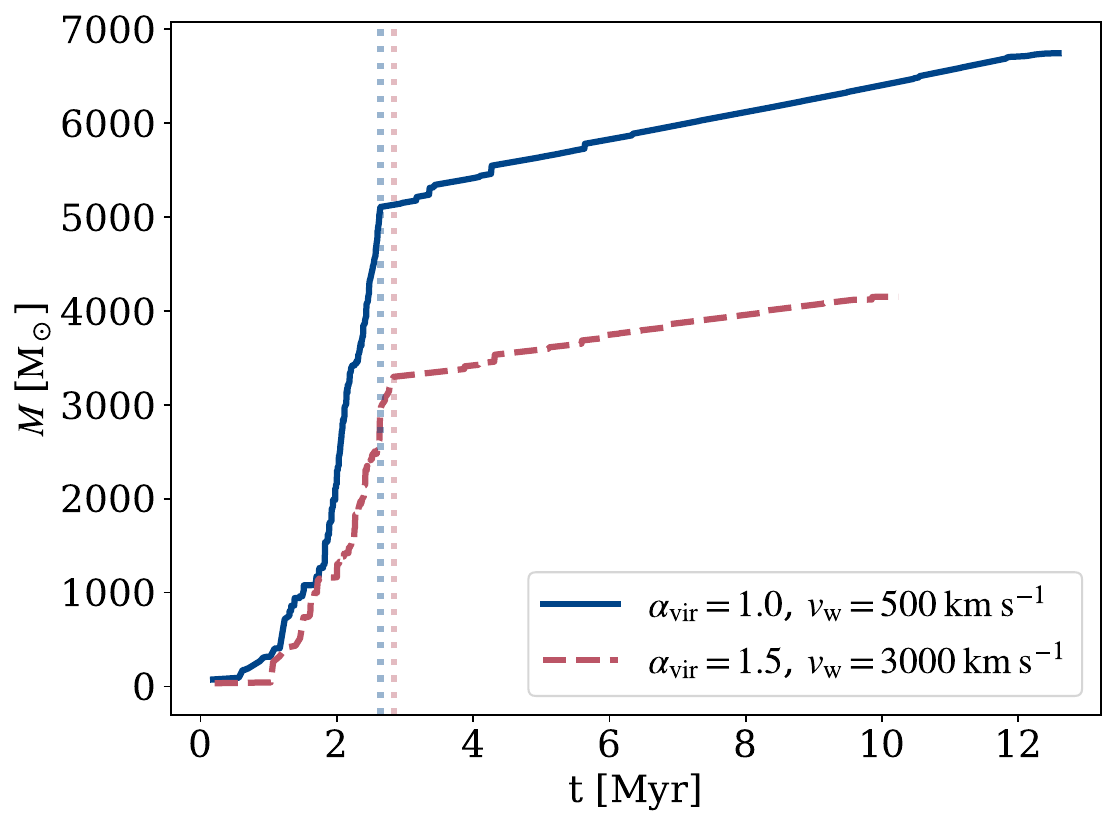}
    \caption{
    Mass evolution of the IMBH and its progenitor VMS, tracked until the gas reservoir is effectively depleted. Two representative models are selected to trace and study the evolution of IMBHs. The vertical dotted lines mark the epoch when the VMS collapses into an IMBH. Note that the mass growth of the VMS is significantly more rapid than that of the IMBH. During the IMBH phase, growth is driven by Edington-limited gas accretion and TDEs, the latter causing occasional step-like increases in mass.
    See Section \ref{subsec:after VMS} for more information.
    }
    \label{VMS_IMBH_evolution}
\end{figure}

\begin{figure*}[t]
    \centering
    \includegraphics[width=\textwidth]{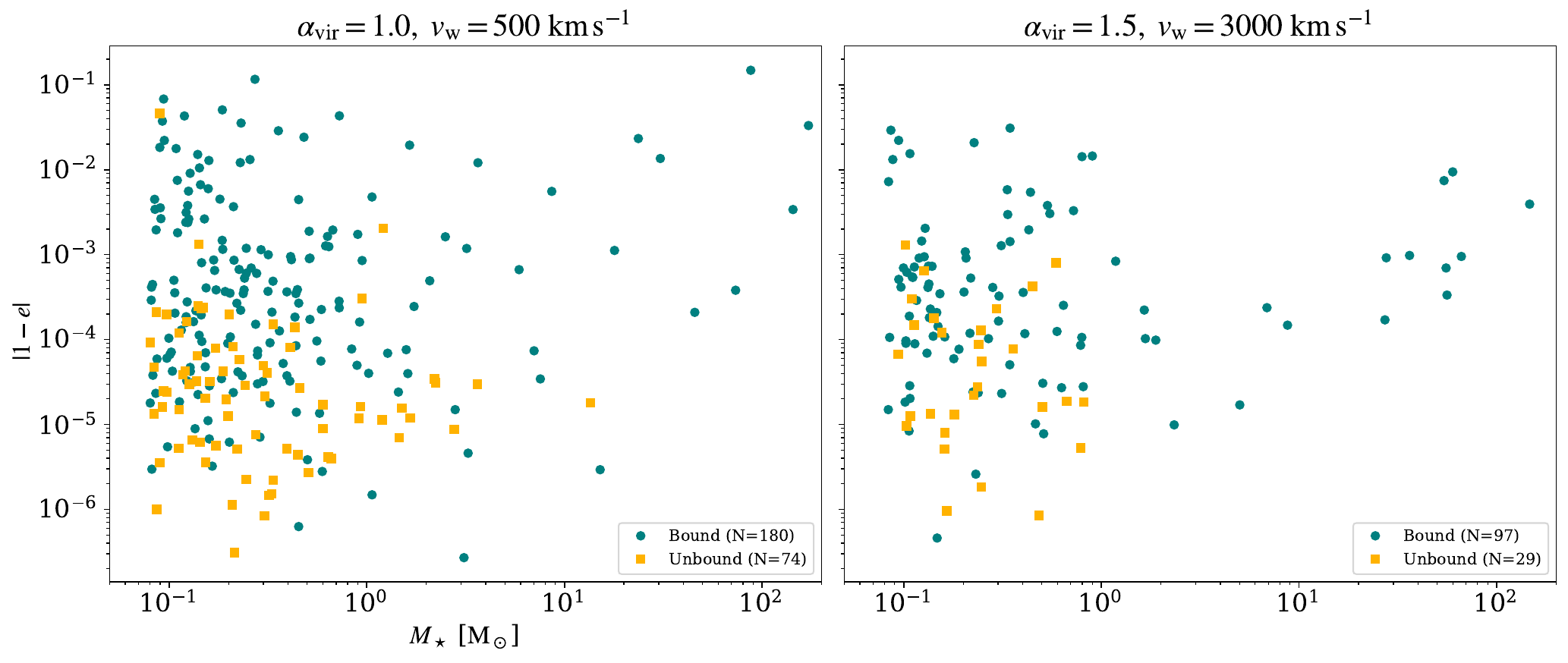}
    \caption{
    Eccentricity versus stellar mass for stars disrupted by the IMBHs in two representative models: $\alpha_\mathrm{vir} = 1.0$, $v_\mathrm{w} = 500\;\mathrm{km\;s^{-1}}$ (\textit{left}; with more massive initial IMBH, $5108\;\mathrm{M_\odot}$) and $\alpha_\mathrm{vir} = 1.5$, $v_\mathrm{w} = 3000\;\mathrm{km\;s^{-1}}$ (\textit{right}; with less massive initial IMBH, $3299\;\mathrm{M_\odot}$).
    The majority of TDEs ($71\;\%$ and $77\;\%$, respectively) originate from bound orbits in both simulations.
    For the more massive IMBH (\textit{left}), one unbound event involves disrupted stars more massive than $8\;\mathrm{M_\odot}$. For the less massive IMBH (\textit{right}), no stars more massive than $8\;\mathrm{M_\odot}$ are disrupted on unbound orbits.
    See Section \ref{subsec:after VMS} for more information.
    }
    \label{fig:TDE_info}
    \vspace{2mm}
\end{figure*}

Figure \ref{VMS_IMBH_evolution} depicts the mass growth history of the IMBH and its progenitor VMS.
In the model with $\alpha_\mathrm{vir} = 1.0$ and $v_\mathrm{w} = 500\;\mathrm{km\;s^{-1}}$, 2543 stellar mergers occur, leading to the formation of a VMS with $M_\mathrm{VMS} = 5108\;\mathrm{M_\odot}$.
In the second run with $\alpha_\mathrm{vir} = 1.5$ and $v_\mathrm{w} = 3000\;\mathrm{km\;s^{-1}}$, the VMS undergoes 1535 stellar mergers, reaching a final mass of $3299\;\mathrm{M_\odot}$.
While the VMS phase is characterized by rapid growth driven by successive stellar mergers, the subsequent IMBH growth is significantly milder, despite having two feeding channels: gas accretion and TDEs.
As defined in Section \ref{subsubsec:TDA} (and also in Section \ref{subsec:merger}), a merger or TDE occurs when the distance between two objects is smaller than the sum of their stellar radii or the tidal radius, respectively.
Since the physical radius of a VMS is substantially larger than the tidal radius, Eq.(\ref{eq_r_TDE}), the VMS grows far more rapidly than the IMBH does.
In the run with a more massive initial IMBH ($\alpha_\mathrm{vir} = 1.0$ and $v_\mathrm{w} = 500\;\mathrm{km\;s^{-1}}$), the $5108\;\mathrm{M_\odot}$ seed grows to $6747\;\mathrm{M_\odot}$ within $10\;\mathrm{Myr}$, corresponding to a mean mass accretion rate of $1.64\times10^{-4}\;\mathrm{M_\odot\;yr^{-1}}$.
In the run with a less massive IMBH, the $3299\;\mathrm{M_\odot}$ seed reaches $4153\;\mathrm{M_\odot}$ within $7.4\;\mathrm{Myr}$, yielding a mean rate of $1.15\times10^{-4}\;\mathrm{M_\odot\;yr^{-1}}$.

In both simulations, we find that mass growth is primarily driven by gas accretion.
Because the initial gas cloud and the resulting star cluster are sufficiently dense to retain a cold gas reservoir at the center, the IMBH accretes gas at the Eddington rate for most of the simulation time, until the gas is eventually expelled from the simulation domain by continuous supernova explosions.\footnote{For both models, the simulations were terminated when the remaining gas mass dropped below 2\% of the initial gas mass. At this stage, the BH accretion rates had settled to a nearly constant value (Figure \ref{VMS_IMBH_evolution}).\label{footnote:gas}} 
Additionally, we note that TDEs also make a substantial contribution, accounting for $23\;\%$ and $34\;\%$ of total mass growth in the run with a more massive and less massive initial IMBH, respectively.  

Figure \ref{fig:TDE_info} shows the mass and eccentricity distributions of all TDEs.
Because the central cluster and IMBH are more massive in the $\alpha_\mathrm{vir} = 1.0$ model, the central gravitational potential is deeper, leading to a higher frequency of TDEs.
In both models, the majority of TDEs originate from bound orbits: $71\;\%$ for the $\alpha_\mathrm{vir} = 1.0$, $v_\mathrm{w} = 500\;\mathrm{km\;s^{-1}}$ model and $77\;\%$ for the $\alpha_\mathrm{vir} = 1.5$, $v_\mathrm{w} = 3000\;\mathrm{km\;s^{-1}}$ model (hereafter, values for the latter are given in parentheses).
Although most TDEs originate from low-mass stars, $4.3\;\%$ ($7.9\;\%$) of events involve massive stars with $M_\star > 8\;\mathrm{M_\odot}$.
Notably, these massive stars are disrupted predominantly on bound orbits.
Specifically, for the more massive IMBH ($\alpha_\mathrm{vir} = 1.0$, $v_\mathrm{w} = 500\;\mathrm{km\;s^{-1}}$), only one massive star is disrupted on an unbound orbit, compared to 10 on bound orbits.
For the less massive IMBH ($\alpha_\mathrm{vir} = 1.5$, $v_\mathrm{w} = 3000\;\mathrm{km\;s^{-1}}$), no massive stars are involved in unbound events, whereas 10 massive stars are disrupted on bound orbits.
We attribute this to the efficient mass segregation of massive stars toward the center, where they are preferentially captured into bound orbits that subsequently decay via post-Newtonian energy dissipation (see Appendix \ref{appendix:PN}).
With half of the disrupted stellar mass accreting onto the IMBH and the rest being ejected (Section \ref{subsubsec:TDA}), TDEs contribute a total of $380\;\mathrm{M_\odot}$ ($294\;\mathrm{M_\odot}$) to the IMBH mass.
Since unbound events are rare and typically involve lower-mass stars --- contributing only $6.4\;\%$ ($1.4\;\%$) of the mass accreted via TDEs --- bound events dominate the growth, accounting for $93.6\;\%$ ($98.6\;\%$).

\begin{figure}[t]
    \centering
    \includegraphics[width=\columnwidth]{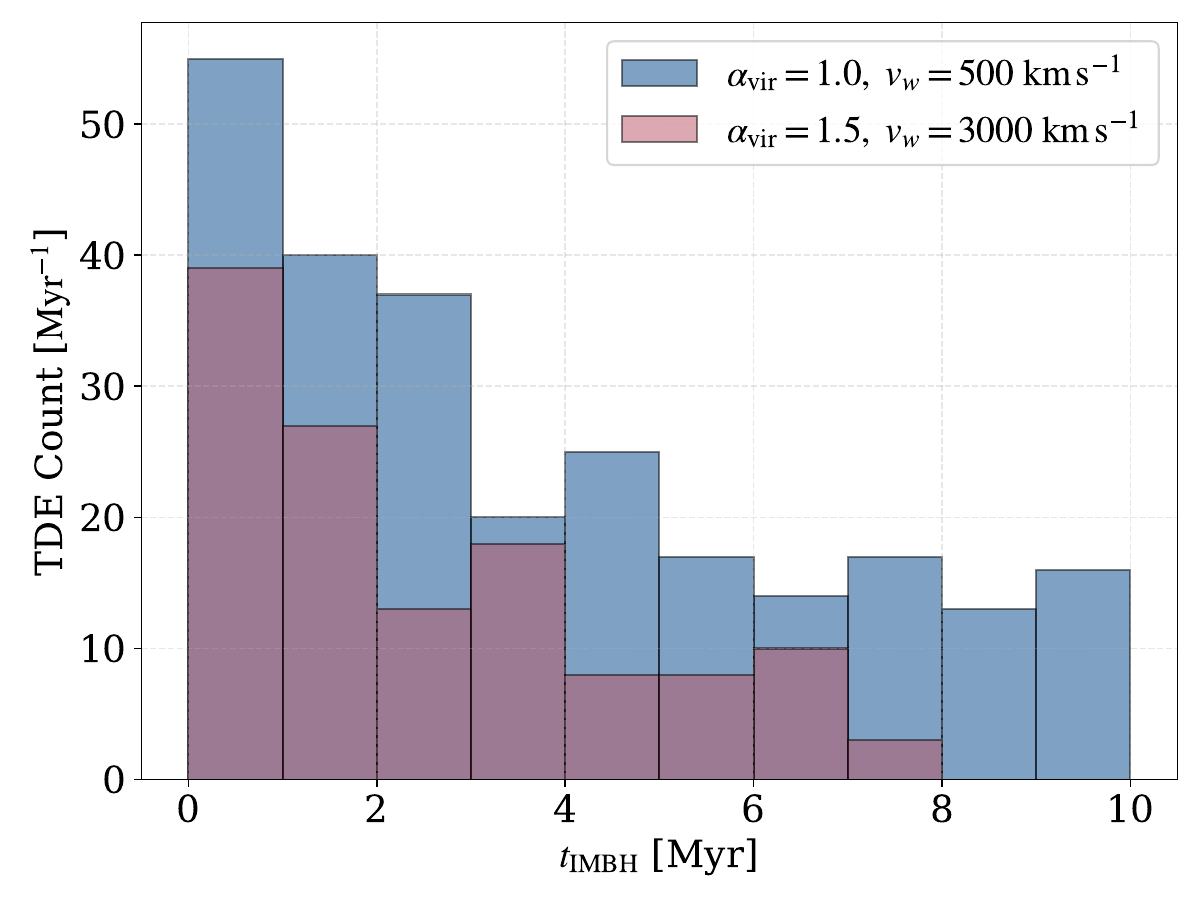}
    \caption{
    TDE histogram.  The time in $x$-axis is measured from the epoch of the IMBH formation for each run. Since the IMBH is more massive in the model with $\alpha_\mathrm{vir} = 1.0$ and $v_\mathrm{w} = 500\;\mathrm{km\;s^{-1}}$, TDEs occur more frequently in this case. The TDE rate peaks immediately after IMBH formation and then gradually decreases in both cases. Note that we ended the $\alpha_\mathrm{vir} = 1.5$ simulation at $t_\mathrm{IMBH} = 7.4\;\mathrm{Myr}$.
    See Section \ref{subsec:after VMS} for more information.
    }
    \label{fig:TDE histogram}
    \vspace{0mm}
\end{figure}

Figure \ref{fig:TDE histogram} illustrates the temporal evolution of the TDE rates following the formation of the IMBH.
Notably, the rate exhibits a sharp peak immediately after the IMBH forms, then gradually declines in both runs.
This initial spike corresponds to the rapid consumption of the pre-existing stellar population within the filled loss cone around the IMBH.
Subsequently, dynamical heating driven by the IMBH and frequent close encounters injects kinetic energy into the surrounding environment, causing the cluster core to expand and the central density to decrease \citep{Shapiro77, Heggie07, Rizzuto23}.
Although quantifying the exact evolution of the core radius in our simulations is challenging due to the complex morphology introduced by merging sub-clusters and ongoing star formation, the decreasing TDE rate serves as a strong indicator of this central density depletion and dynamical relaxation.

\begin{figure*}[t]
    \centering
    \includegraphics[width=1.0\textwidth]{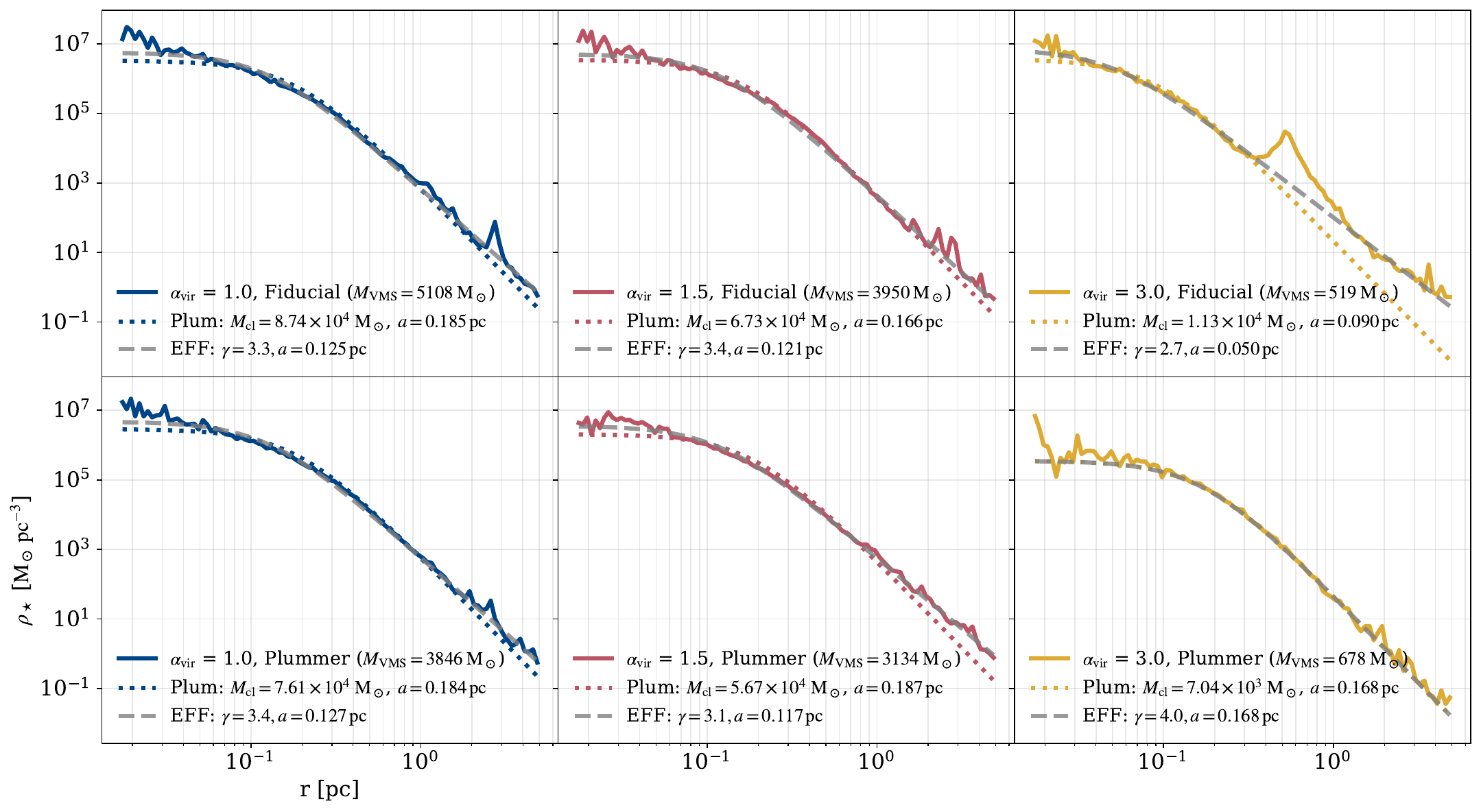}
    \caption{
    Stellar density profiles with best-fit Plummer (\textit{dotted lines}) and EFF models (\textit{gray, dashed lines}) immediately prior to the VMS collapse --- for the fiducial initial gas distribution (\textit{top}; Eq.(\ref{eq:fiducial profile})) and the Plummer initial gas distribution (\textit{bottom}; Eq.(\ref{eq:plummer profile})).
    The best-fit parameters for the Plummer models ($M_\mathrm{cl}$ and $a$), parameters for the EFF models ($\gamma$ and $a$), and the resulting VMS (IMBH progenitor) mass are indicated in each panel.
    By the time the VMS collapses into an IMBH, most simulations show a density structure broadly consistent with a Plummer profile in the intermediate regions, while the EFF profiles provide a superior fit to the outer envelopes.
    A notable exception is the fiducial model with $\alpha_\mathrm{vir} = 3.0$ (\textit{top right}), which is poorly fitted by a Plummer profile due to an ongoing sub-cluster merger.
    For the Plummer model with $\alpha_\mathrm{vir} = 3.0$ (\textit{bottom right}), the best-fit EFF model is consistent with the Plummer model.
    See Section \ref{subsec:plummer} for more information.
    }
    \label{plummer fitting}
    \vspace{3mm}
\end{figure*}

In addition to the central IMBH, our simulations produce a population of stellar-mass compact objects.
In principle, these objects could participate in TDEs, BH-BH mergers, or BH-IMBH mergers.
However, due to the limited simulation timescale, we detected no BH-BH mergers or TDEs involving stellar-mass remnants (see Section \ref{subsec:merger}), nor did these remnants contribute to the evolution of the  IMBH.
The formation and distribution of the stellar-mass BHs are further discussed in Appendix \ref{appendix:stellar-BH}.

\hspace{1mm}

\section{Discussion} \label{sec:discussion}

\subsection{Sensitivity of the Star Cluster Evolution on the Initial Gas Profiles} \label{subsec:plummer}

To model dense star clusters and investigate the formation of VMSs, we initialize turbulent gas clouds following the density profile defined in Eq.(\ref{eq:fiducial profile}).
This profile serves as our fiducial model, selected due to the lack of observational constraints on the precise internal structure of high-redshift cluster progenitors.
To quantify how the initial density distribution affects cluster assembly and runaway collisions, we also test an initial condition following a Plummer profile \citep{Plummer11}:
\begin{equation}
    \rho(r) = \rho_\mathrm{c} \left( 1 + \frac{r^2}{a^2} \right) ^ {-5/2},
    \label{eq:plummer profile}
\end{equation}
where $\rho_\mathrm{c} = 3M_\mathrm{cl}/(4\pi a^3)$.
We adopt parameters $\rho_\mathrm{c} = 2.42\times10^{20}\;\mathrm{g\;cm^{-3}}$ and $a = 5\;\mathrm{pc}$ to ensure the total gas mass within the simulation box matches that of the fiducial models.
The wind velocity for massive stars is fixed to $500\;\mathrm{km\;s^{-1}}$ and we run 3 simulations with different initial turbulence of $\alpha_\mathrm{vir} = 1.0$, $1.5$, and $3.0$.

Figure \ref{plummer fitting} displays the stellar density profiles of our simulations immediately prior to the VMS collapse, overlaid with their best-fit Plummer and Elson-Fall-Freeman (EFF) profiles \citep{Elson87}.
The EFF profile is defined as
\begin{equation}
    \rho(r) = \rho_0\left(1 + \frac{r^2}{a^2} \right)^{-(\gamma+1)/2},
\end{equation}
where the Plummer profile is a special case with $\gamma = 4$.
In general, the simulated clusters are well described by Plummer profiles except in the innermost and outermost regions.
However, for the fiducial model with $\alpha_\mathrm{vir} = 3.0$, the density profile is poorly matched by the Plummer model because a cluster merger occurred immediately prior to the VMS collapse, leaving the main cluster in a dynamically unrelaxed state.
For all models, EFF profiles provide a superior fit in the outer regions, with a best-fit power-law index of $\gamma\sim3$.
The core regions remain difficult to fit with these standard analytic models because the simulated clusters develop central density cusps --- the primary driver of runaway stellar collisions.
These cusps drive the formation of VMSs with $M_\mathrm{VMS} > 500\;\mathrm{M_\odot}$ regardless of the initial gas distribution.
Notably, a denser initial gas profile does not necessarily yield a denser star cluster; the resulting massive clusters exhibit similar structural properties and VMS masses across different initial density models.

While our simulations demonstrate that different initial gas profiles yield star clusters consistent with Plummer models, confirming the exact gas distribution and resulting stellar distributions at high redshift remains challenging due to their compact spatial scales.
Furthermore, although we explored a range of initial turbulence levels from the virial equilibrium ($\alpha_\mathrm{vir} = 1.0$) to a highly turbulent state ($\alpha_\mathrm{vir} = 3.0$), the high-redshift environment can be even more dynamic, driven by frequent structural assemblies and intense gas inflows into the central region of galaxies.
For instance, \citet{Lahen25_merger} simulated the merger of two gas-rich, low-metallicity dwarf galaxies and found the formation of dense star clusters reaching surface densities of $\sim 7\times 10^4\;\mathrm{M_\odot\;pc^{-2}}$ and hosting a VMS of $\sim1000\;\mathrm{M_\odot}$.
To fully capture the realistic configuration of these chaotic high-redshift environments, however, cosmological simulations with individual-star resolution are necessary (see Section \ref{subsec:future}).

\subsection{Extrapolated Mass Evolution for IMBH} \label{subsec:IMBH extrapolation}

\begin{figure}[t]
    \centering
    \includegraphics[width=\columnwidth]{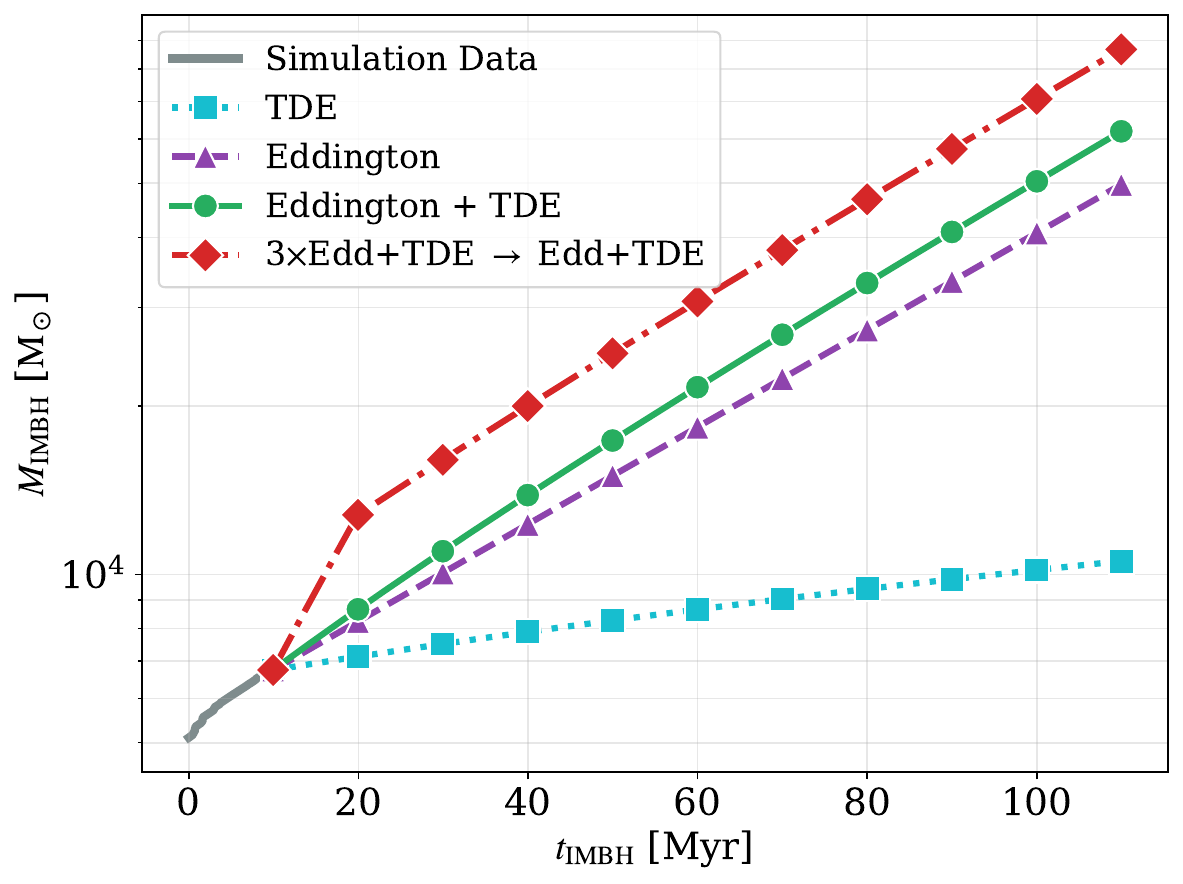}
    \caption{
    Projected mass growth of the IMBH over the next $100\;\mathrm{Myr}$.
    The gray line shows the simulated mass evolution of the most massive IMBH formed in our suite (reaching $M_\mathrm{BH} = 6747\;\mathrm{M_\odot}$ from the run with $\alpha_\mathrm{vir} = 1.0$ and $v_\mathrm{w} = 500\;\mathrm{km\;s^{-1}}$) up to the end of the run. For the extrapolation, we assume a constant TDE accretion rate of $3.82\times10^{-5}\;\mathrm{M_\odot\;yr^{-1}}$.
    The cyan dotted line (\textit{squares}) shows growth driven by TDEs only; the purple dashed line (\textit{triangles}) assumes gas accretion at the Eddington rate only; and the green solid line (\textit{circles}) combines both TDE and Eddington accretion.
    The red dash-dotted line (\textit{diamonds}) represents a super-Eddington scenario: it combines the constant TDE rate with gas accretion at three times the Eddington rate for the first $10\;\mathrm{Myr}$, followed by standard Eddington accretion for the remaining $90\;\mathrm{Myr}$.
    See Section \ref{subsec:IMBH extrapolation} for more information.
    }
    \label{fig:extrapolation}
\end{figure}

Due to the high computational cost of resolving individual stars, our simulations employ small, isolated volumes without cosmological gas inflows.
Consequently, gas accretion onto the IMBH is quenched after $\sim 10\;\mathrm{Myr}$ once supernova feedback expels the initial gas reservoir.
However, high-redshift star clusters likely reside in gas-rich, dynamically active environments where galaxy mergers and cold streams replenish the gas supply \citep{DiMatteo05, Dekel09}.
To estimate the further evolution of these IMBHs, we analytically extrapolate their growth assuming a gas-rich environment.
While gas accretion at the Eddington rate remains the dominant growth mechanism in our model (see Section \ref{subsec:after VMS}), we also include a constant TDE-driven accretion rate for our extrapolation as well.
In our isolated simulations, the TDE-driven mass accretion rate naturally declines after star formation quenches, as the population of massive stars decreases due to their short lifetime.
In reality, continuous gas inflows could sustain star formation, and cluster mergers could increase the central stellar density, maintaining --- or potentially even boosting --- the TDE rate.    

Figure \ref{fig:extrapolation} illustrates the estimated evolution of the IMBH over the next $100\;\mathrm{Myr}$, starting from the most massive IMBH ($M_\mathrm{BH} = 6747\;\mathrm{M_\odot}$) found in our simulation suite (in the run with $\alpha_\mathrm{vir} = 1.0$ and $v_\mathrm{w} = 500\;\mathrm{km\;s^{-1}}$).
We adopt a constant tidal disruption accretion rate of $3.82 \times 10^{-5}\;\mathrm{M_\odot\;yr^{-1}}$ derived from the $380\;\mathrm{M_\odot}$ accreted via 254 TDEs over $9.96\;\mathrm{Myr}$ (see Section \ref{subsec:after VMS}).
We compare four distinct growth scenarios: (1) TDE accretion only; (2) gas accretion only at the Eddington rate; (3) combined TDE and Eddington accretion; and (4) combined TDE and super-Eddington accretion ($3\times \dot{M}_\mathrm{Edd}$).
For the last case, we assume the super-Eddington accretion lasts only $10\;\mathrm{Myr}$, followed by standard Eddington accretion for the remaining $90\;\mathrm{Myr}$, as high accretion rates likely trigger strong AGN feedback that suppresses further inflow \citep{Wu25}.
In the TDE-only scenario, the IMBH of $6747\;\mathrm{M_\odot}$ is estimated to grow to $10562\;\mathrm{M_\odot}$ within $100\;\mathrm{Myr}$.
With Eddington accretion alone, the IMBH reaches a mass of $49742\;\mathrm{M_\odot}$.
When TDEs are coupled with Eddington accretion, the black hole grows more rapidly, reaching $61912\;\mathrm{M_\odot}$.
Finally, in the super-Eddington scenario, the IMBH evolves to $86731\;\mathrm{M_\odot}$, demonstrating that even short phases of super-Eddington accretion can dramatically accelerate IMBH evolution.
With such high masses, these objects are capable of serving as a seed for a SMBH in the early Universe.

Furthermore, recent JWST observations have revealed a population of high-redshift sources, which is called the Little Red Dots \citep[LRDs; ][]{Kocevski23, Harikane23, Matthee24, Kokorev24}.
While the exact origin of LRDs is not fully understood, the stellar interpretation suggests they are intensely star-forming dusty galaxies with extreme central stellar densities --- the conditions where runaway mergers are expected to occur \citep{Guia24, Akins25, Rantala26_LRD}.
We argue that IMBHs could naturally form and rapidly evolve within such environments, potentially explaining the compact nature of LRDs under this scenario \citep{Pacucci25, Escala25}.

While we simply assumed constant TDE rates, dynamic processes such as cluster mergers and star formation bursts driven by intense gas inflows could significantly enhance the TDE rates over time.
Furthermore, the merger of massive clusters could lead to the coalescence of their central IMBHs, providing an additional growth channel.
Consequently, our current estimates likely represent a conservative lower limit on the evolution of IMBHs.
While AGN feedback by SMBHs might suppress the gas inflow \citep{DiMatteo05, Fabian12}, we argue that IMBHs can sustain the gas accretion at the Eddington rate, provided they reside in galactic centers fed by steady, dense gas inflows that overwhelm the thermal feedback from relatively low-mass IMBHs \citep{Inayoshi16, Takeo20}.

\subsection{Future Work} \label{subsec:future}

We plan to extend our simulation framework to study the formation and evolution of SMBH seeds in more realistic contexts. 
The specific objectives are as follows:

\begin{itemize}
    \item 
    \textit{Running fully cosmological zoom-in simulations.}
    In this study, we performed simulations using isolated gas clouds with initialized turbulence to approximate the physical conditions of the high-redshift Universe. 
    However, isolated setups cannot fully capture the complex assembly history of early galaxies, such as continuous cold gas accretion, external tidal fields, and hierarchical mergers.
    In future studies, we will employ cosmological zoom-in simulations to self-consistently include these environmental factors.
    This will allow us to capture the formation of specific dense environments --- such as proto-globular clusters resembling the Cosmic Gems clusters \citep{Adamo24, Bradley25, Vanzella25} or proto-nuclear star clusters --- that naturally foster the extreme densities required to trigger runaway stellar collisions.
    
    \item 
    \textit{Implementing Pop III physics and radiative transfer.}
    To accurately identify the formation of dense stellar systems at high redshift, it is necessary to incorporate appropriate Pop III physics.
    Furthermore, accurate radiative feedback is critical, as the intense radiation fields from massive stars can heat the gas and alter subsequent star formation.
    While the present study utilized a simplified HII region feedback model (see Section \ref{subsubsec:HII}), future work requires a more rigorous treatment, such as adaptive ray-tracing radiative transfer.
    Adopting the AEOS physics module \citep{Brauer25} can be a good option.
    Specifically, we plan to couple AEOS with the stellar evolution code \texttt{SEVN}, ensuring that individual stellar evolution, radiation, and feedback are treated self-consistently within the \texttt{Enzo-Abyss} framework.
    We will be able to track detailed stellar yields, specifically from VMS winds and TDEs, by following individual metal species.
    Ultimately, we aim to investigate nitrogen enrichment in dense stellar environments and compare our predictions with observations of high-z, nitrogen-enhanced galaxies such as GN-z11 \citep{Oesch16, Cameron23} and CEERS-1019 \citep{Larson23, Marques-Chaves24}.
\end{itemize}


\section{Summary and Conclusion} \label{sec:conclusion}

We introduce an updated version of \texttt{Enzo-Abyss}, a hydrodynamics code employing a direct $N$-body method to accurately resolve gravitational dynamics between stars.
To follow individual stars down to a mass of $0.08\;\mathrm{M_\odot}$, we couple the code with the external libraries, such as \texttt{SDAR} for accurate few-body dynamics and \texttt{SEVN} for individual stellar evolution.
Using this framework, we model centrally dense gas clouds with a total mass of $1.75 \times 10^5\;\mathrm{M_\odot}$ and varying levels of initial turbulence ($\alpha_\mathrm{vir} = 1.0$, $1.5$, and $3.0$) and stellar wind feedback strength to investigate the formation of massive star clusters that resemble the compact systems observed at high redshift. 
Our key findings are as follows:
\begin{itemize}
    \item 
    Massive star clusters with masses ranging from $2.7\times10^4\;\mathrm{M_\odot}$ to $8.8 \times 10^4\;\mathrm{M_\odot}$ form within $4\;\mathrm{Myr}$ of the first star formation event (Section \ref{subsec:formation of massive star clusters}).
    Regardless of the initial gas density profile, the resulting clusters are well fitted by Plummer profiles (Section \ref{subsec:plummer}).
    Their half-mass radii are smaller than $1\;\mathrm{pc}$, and their surface densities lie in the range of $5\times10^3\;\mathrm{M_\odot\;pc^{-2}} \leq \Sigma \leq 3\times10^5\;\mathrm{M_\odot\;pc^{-2}}$.
    These structural properties are directly comparable to the high-z star clusters identified in recent JWST observations.
    
    \item 
    In the dense cores of massive star clusters, VMSs can form via runaway collisions if central densities are consistently maintained above $\sim 10^6\;\mathrm{M_\odot \; pc^{-3}}$, a threshold proposed by \citet{Ardi08} and \citet{Fujii24}.
    In our simulations, models with $\alpha_\mathrm{vir} = 1.0$ and $1.5$ satisfy this criterion, forming VMSs with masses exceeding $3000\;\mathrm{M_\odot}$ (Section \ref{subsec:VMS evolution}).
    In contrast, the $\alpha_\mathrm{vir} = 3.0$ models barely reach this density threshold, doing so only in the innermost regions.
    Consequently, significantly lighter VMSs form in these cases ($< 600\;\mathrm{M_\odot}$.)
    Nevertheless, even these lower-mass VMSs are sufficiently massive to avoid the PISN gap and directly collapse into IMBHs.
    
    \item
    After the VMSs collapse into IMBHs, IMBHs evolve via gas accretion and TDEs.
    Supported by the cold gas reservoir within the massive star clusters, IMBHs accrete gas at the Eddington rate until successive supernova explosions expel most of the remaining gas (Section \ref{subsec:after VMS}).
    In one simulation with the $5108\;\mathrm{M_\odot}$ IMBH, we find an average accretion rate of $1.64\times10^{-4}\;\mathrm{M_\odot \; yr^{-1}}$ over a period of $10\;\mathrm{Myr}$.
    We identify $254$ TDEs, which account for $23\;\%$ of the total accreted mass;
    notably, $71\;\%$ of these events originate from bound orbits.
    
    \item 
    We extrapolate the IMBH evolution to another $100\;\mathrm{Myr}$ by combining a constant TDE rate derived from our simulations with continuous gas accretion at the Eddington rate, assuming a gas-rich environment with frequent structural assemblies and steady gas flows into the young NSCs at high redshift (Section \ref{subsec:IMBH extrapolation}).
    Under these conditions, the $6747\;\mathrm{M_\odot}$ IMBH reaches $61912\;\mathrm{M_\odot}$ within $100\;\mathrm{Myr}$.
    If we further assume that a super-Eddington accretion rate ($3\times \dot{M}_\mathrm{Edd}$) is sustained for the initial $10\;\mathrm{Myr}$, the final mass increases to $86731\;\mathrm{M_\odot}$.
    We argue that IMBHs formed in dense NSCs can rapidly grow via gas accretion and TDEs, offering a plausible pathway for seeding high-redshift SMBHs.
\end{itemize}

Using a single self-consistent framework  that incorporates all relevant physical processes, we successfully demonstrate how VMSs form in the cores of massive star clusters, collapse into IMBHs, and subsequently evolve via gas accretion and TDEs.
However, the detailed gas distribution and the precise structure of the cluster cores remain difficult to constrain due to the limitation of spatial resolution of current observations.
Future work will address this by integrating our direct $N$-body algorithms into high-resolution cosmological zoom-in simulations.
This approach will enable more realistic initial conditions and allow us to investigate IMBH formation within a fully cosmological context.

\section*{Acknowledgments}

We would like to thank Harley Katz for his kind and insightful comments on the earlier version of this manuscript.
J.-H.K.’s work was supported by the National Research Foundation of Korea (NRF) grant funded by the Korea government (MSIT) (No. 2022M3K3A1093827 and No. 2023R1A2C1003244).
His work was also supported by the Korea Institute of Science and Technology Information (KISTI) through supercomputing resources, including technical support for code parallelization, under grants KSC-2024-CRE-0232.
His work was further supported by the GlobalLAMP Program of the NRF grant funded by the Ministry of Education (No. RS-2023-00301976). 
O.-K. K. was supported by the KISTI under the institutional R\&D project (K26L1M2C3).

\appendix

\section{Orbit Averaged Post-Newtonian Corrections} \label{appendix:PN}

Compact binary systems lose orbital energy through the emission of gravitational waves (GWs), causing their orbits to shrink and the components to eventually coalesce.
To account for this general relativistic effect, we apply post-Newtonian corrections to the orbital parameters.
Following the orbital integration step in \texttt{SDAR}, we apply orbit-averaged post-Newtonian corrections to the semi-major axis ($a$) and eccentricity ($e$):
\begin{equation}
    \Delta a = -\frac{64}{5}\frac{G^3 m_1 m_2 (m_1 + m_2)}{a^3 c^5} \frac{1 + \frac{73}{24}e^2 + \frac{37}{96}e^4}{(1-e^2)^{7/2}} \Delta t
\end{equation}
\begin{equation}
    \Delta e = -\frac{304}{15}\frac{G^3 m_1 m_2 (m_1 + m_2)}{a^4 c^5} \frac{e + \frac{121}{304}e^3}{(1-e^2)^{5/2}} \Delta t,
\end{equation}
where $m_1$ and $m_2$ are the masses of the binary components and $\Delta t$ is the integration timestep \citep{Peters64}.

\section{Stellar-mass BH distribution} \label{appendix:stellar-BH}

\begin{figure}[t]
    \centering
    \includegraphics[width=\columnwidth]{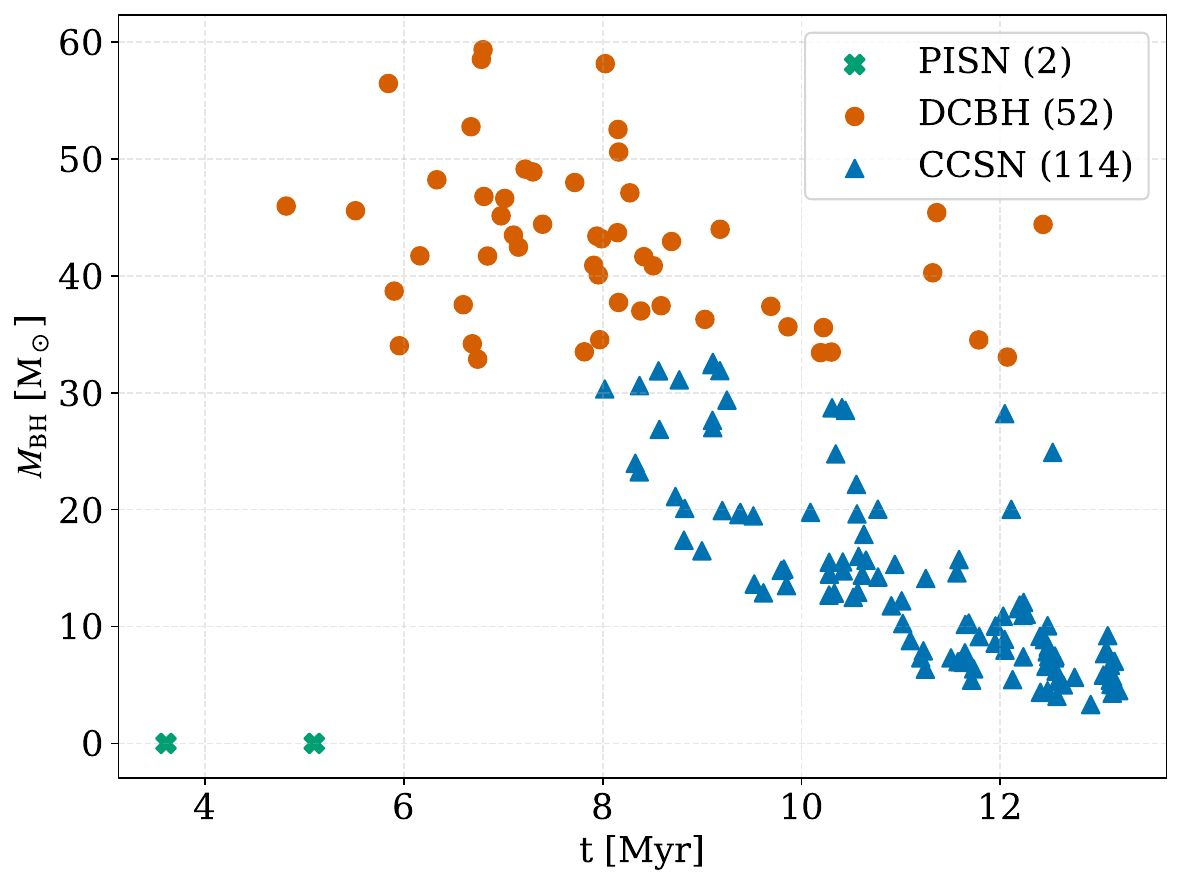}
    \caption{
    Distribution of stellar-mass BH remnant masses as a function of their formation time. The markers distinguish the remnant formation channels: PISN (\textit{green cross}), DCBH (\textit{orange circles}), and CCSN (\textit{blue triangles}). The numbers in the legend indicate the total count of events for each channel. DCBH events (failed supernovae) dominate the early epoch ($t \lesssim 8\;\mathrm{Myr}$) and produce more massive remnants ($M_\mathrm{BH} \gtrsim 30\;\mathrm{M_\odot}$), whereas CCSN events occur later and result in lower-mass BHs. The two PISN events at $t = 3.6\;\mathrm{Myr}$ and $5.1\;\mathrm{Myr}$ results in the total disruption of the star, leaving no remnant.
    See Appendix \ref{appendix:stellar-BH} for more information.
    }
    \label{remnant formation}
\end{figure}

\begin{figure}[t]
    \centering
    \includegraphics[width=\columnwidth]{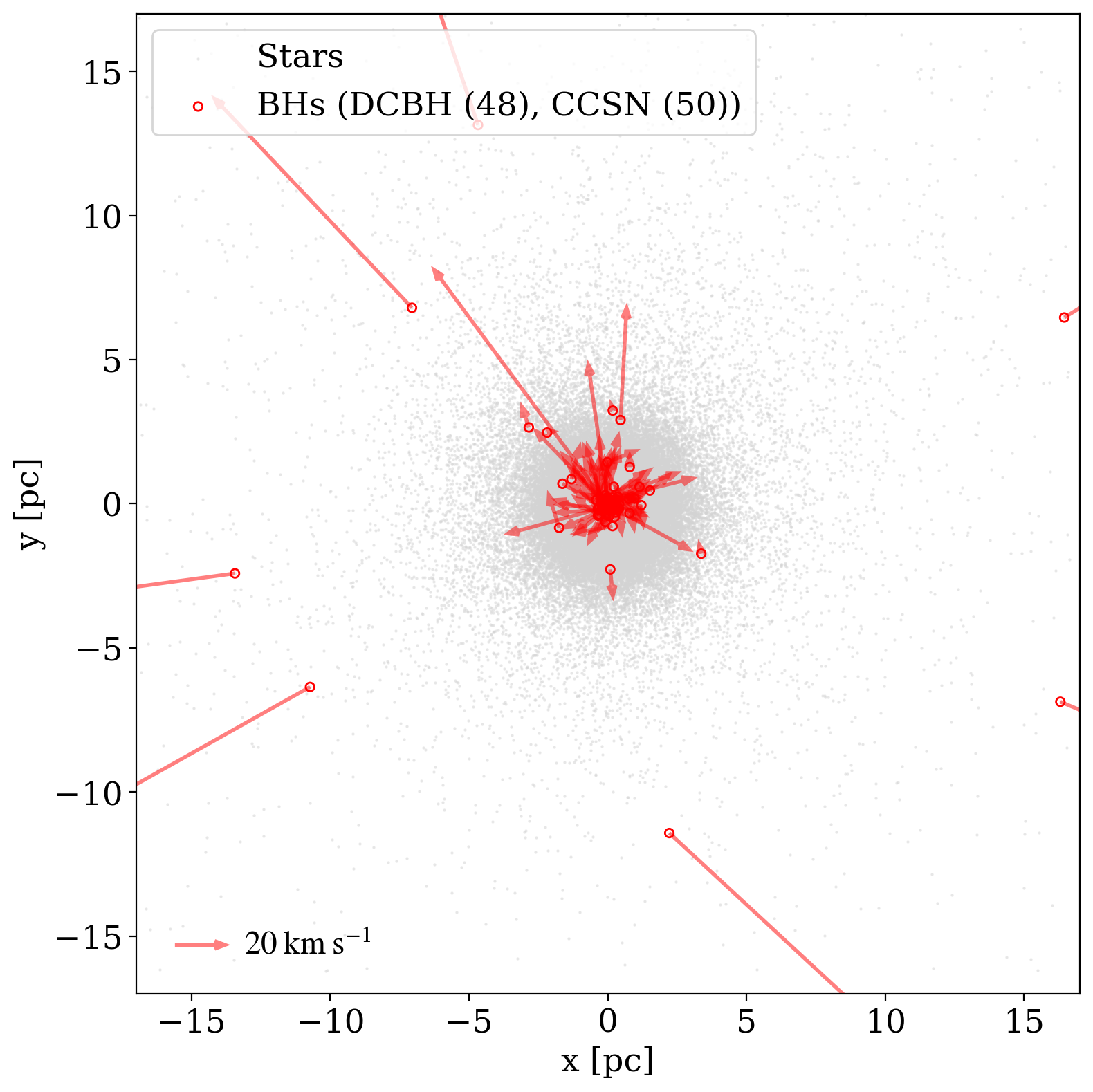}
    \caption{
    Projected spatial distribution of stars (\textit{grey dots}) and stellar-mass BHs (\textit{red circles}) in the central region of the cluster. The arrows indicate the projected velocity vectors of the BHs, with a reference scale of $20\;\mathrm{km\;s^{-1}}$ shown in the bottom left. The BH population consists of 48 DCBHs and 50 CCSN remnants. While the majority of remnants remain gravitationally bound in the cluster core, a subset of BHs exhibits high velocities and large radial offsets, driven by the natal kicks received during CCSNe.
    See Appendix \ref{appendix:stellar-BH} for more information.
    }
    \label{fig:BH distribution}
\end{figure}

Figure \ref{remnant formation} shows the mass and creation time of individual stellar-mass BHs formed after the IMBH formation in the model with $\alpha_\mathrm{vir} = 1.0$ and $v_\mathrm{w} = 500\;\mathrm{km\;s^{-1}}$.
We identify three distinct outcomes at the end of the stellar life cycle:
PISN explosion, direct collapse BH (DCBH) formation, and BH formation accompanied by a CCSN.
While lighter stars would evolve into NSs or WDs, we do not observe them because the simulation duration is shorter than their lifetimes.

The first SN event occurs $3.6\;\mathrm{Myr}$ after the first star formation event.
The progenitor star, formed via multiple stellar mergers, had a ZAMS mass of $200\;\mathrm{M_\odot}$ and ended its life as a PISN without leaving a remnant.
At $t = 4.8\;\mathrm{Myr}$, the first stellar-mass BH ($46\;\mathrm{M_\odot}$) formed from a progenitor with an initial mass of $106\;\mathrm{M_\odot}$.
With a fallback fraction of $1.0$, this star produced no supernova and collapsed directly into a BH.
The first CCSN was delayed until $8.0\;\mathrm{Myr}$ after the first star formation.
Thereafter, a series of 113 CCSNe occurred over the next $4.6\;\mathrm{Myr}$, effectively clearing the gas from the simulation volume.

For BHs formed via CCSN events, we apply natal kicks calculated in \texttt{SEVN} following the method of \citet{Giacobbo20}.
The magnitude of the kick is given by
\begin{equation}
    v_\mathrm{kick} = f_\mathrm{H05}\frac{m_\mathrm{ej}}{\langle m_\mathrm{ej} \rangle}\frac{\langle m_\mathrm{NS} \rangle}{m_\mathrm{rem}},
\end{equation}
where $f_\mathrm{H05}$ is drawn randomly from a Maxwellian distribution with $\sigma = 265\;\mathrm{km\;s^{-1}}$ \citep{Hobbs05}, $m_\mathrm{ej}$ is the ejecta mass,  $m_\mathrm{rem}$ is the remnant mass, and $\langle m_\mathrm{ej} \rangle = 9\;\mathrm{M_\odot}$ and $\langle m_\mathrm{NS} \rangle = 1.2\;\mathrm{M_\odot}$ are the average ejecta and NS masses, respectively (see \citet{Giacobbo20} for details).
Consequently, massive BHs with high fallback fractions (low ejecta mass) receive negligible kicks.
In contrast, lower-mass BHs with low fallback fraction receive significantly larger kicks ($v_\mathrm{kick} > 100\;\mathrm{km\;s^{-1}}$), often sufficient to escape the cluster.
We note that recent reassessments of pulsar distances \citep[e.g.,][]{Verbunt17, Igoshev20, Disberg25} suggest that the \citet{Hobbs05} analysis may be subject to systematic errors, favoring significantly lower kick velocities. 
If lower natal kicks were adopted, the retention fraction of lower-mass BHs in our simulations would likely increase.
However, to remain consistent with the current implementation of \texttt{SEVN} and \citet{Giacobbo20}, we retain the \citet{Hobbs05} baseline for this study.

Figure \ref{fig:BH distribution} shows the spatial distribution of stellar BHs with their velocities at the final snapshot ($t = 13.22\;\mathrm{Myr}$) in the model with $\alpha_\mathrm{vir} = 1.0$ and $v_\mathrm{w} = 500\;\mathrm{km\;s^{-1}}$.
Consistent with mass segregation, the majority of BHs reside in the central region.
However, we also observe a number of escaping BHs, in addition to 68 that have already completely exited the simulation box.
Of the 114 BHs formed via CCSN, only 50 remain within the simulation volume.
In contrast, only 4 DCBHs have exited the domain; although DCBHs do not receive natal kicks, they can still be ejected via dynamical interactions, such as three-body encounters or gravitational slingshots involving the central IMBH.

\bibliography{main}{}
\bibliographystyle{aasjournal}

\end{document}